\documentstyle[epsf]{mn} \input epsf
\newcommand{\eemean}{\mbox{$\langle e^2 \rangle$}}
\newcommand{\iimean}{\mbox{$\langle i^2 \rangle$}}
\newcommand{\erms}{\mbox{$\langle e^2 \rangle^{1/2}$}}
\newcommand{\irms}{\mbox{$\langle i^2 \rangle^{1/2}$}}
\newcommand{\ermsj}{\mbox{$\langle e_*^2 \rangle^{1/2}$}}
\newcommand{\irmsj}{\mbox{$\langle i_*^2 \rangle^{1/2}$}}
\newcommand{\eq}{\begin{equation}}
\newcommand{\eqend}{\end{equation}}
\newcommand{\disp}{\displaystyle}
\begin{document}
\title[Velocity dispersion in disc potentials]
    {Evolution of the Velocity Dispersion of Self-Gravitating 
     Particles in Disc Potentials}

\author[K. Shiidsuka and S. Ida]
    {Kouji Shiidsuka
  and Shigeru Ida \\
     Department of Earth and Planetary Sciences, 
     Faculty of Science, 
     Tokyo Institute of Technology,
     Tokyo 152-8551, Japan \\
        {\rm E-mail: kshiidsu@geo.titech.ac.jp}}
\maketitle  

\begin{abstract}
The ratio of the vertical velocity dispersion to
radial one ($\sigma_z / \sigma_R$) of 
self-gravitating bodies in various disc potentials is investigated
through two different numerical methods
(statistical compilation of two-body
encounters and $N$-body simulations).
 The velocity dispersion generated by two-body relaxation is considered.
The ratio is given as a function of 
 a disc potential parameter,
$\kappa/\Omega$, where $\kappa$ and $\Omega$ are
the epicycle and circular frequencies
(the parameters $\kappa/\Omega=1$ and 2 correspond to
Kepler rotation and solid-body rotation). 
For $1 \leq \kappa/\Omega \la 1.5$, 
the velocity dispersion increases
keeping some anisotropy ($\sigma_z / \sigma_R \sim 0.5$-$0.7$) 
if the amplitude
of radial excursion is larger than tidal radius, while
$\sigma_z / \sigma_R \ll 1$ for smaller amplitude.
On the other hand, for $1.5 \la \kappa/\Omega \leq 2.0$,
we found isotropic state ($\sigma_z / \sigma_R \simeq 1$)
in the intermediate velocity regime, while anisotropic
state ($\sigma_z / \sigma_R < 1$) still exists for higher 
and lower velocity regimes.
The range of the intermediate velocity regime expands
with $\kappa/\Omega$.
In the limit of solid-body rotation, the regime
covers all over the velocity space.
Thus, the velocity dispersion generally has two 
different anisotropic states for each disc potential 
($1 \leq \kappa/\Omega < 2$)
and one isotropic state for $1.5 \la \kappa/\Omega < 2$
where the individual states correspond to different amplitude of velocity disper
sion,
while in the limit of solid-body rotation ($\kappa/\Omega = 2.0$), 
entire velocity space is covered by the isotropic state.

\end{abstract}


\section{INTRODUCTION}  

Gravitational interactions between bodies in a disc potential tend to 
increase velocity dispersion of the bodies as well as diffuse the 
bodies radially.
The increased random energy is transferred from the potential energy 
through the radial diffusion. 
This process is called 'disc heating' for stars and molecular clouds 
in the Galactic gravitational field 
(e.g. Spitzer \& Schwarzshild 1953; Binney \& Tremaine 1987; Lacey 1991),
and 'viscous stirring' for planetesimals in the protoplanetary disc or 
ring particles around planets (e.g. Stewart \& Wetherill 1988; Ida 1990).
Hereafter we use 'disc heating'.
In general, disc heating results in anisotropic velocity dispersion.
The radial and the vertical components of the velocity dispersion 
($\sigma_{R}$ and $\sigma_{z}$) evolve 
with keeping a certain 'equilibrium' ratio, 
which is not generally unity. 

Ida \& Makino (1992) showed through $N$-body simulations 
that $\sigma_{z}/\sigma_{R} \simeq 0.45$ for self-gravitating planetesimals in the 
solar (Keplerian) potential.
Numerical simulations of disc stars perturbed by massive melocular clouds in the solar 
neighbourhood showed $\sigma_z /\sigma_R \simeq 0.6$
(Villumsen 1985; Kokubo \& Ida 1992). 
Observations of stars in the solar neighbourhood show consistent 
anisotropy (e.g., Wielen 1977; Chen, Asiain, Figueras \& Torra 1997). 

 In our galaxy, collective effects such as transient density waves 
would play an important role in velocity dispersion of stars in the solar neighbourhood 
(e.g., Barbanis \& Woltjer 1967; Binney \& Lacey 1988; Jenkins \& Binney 1990). 
Nevertheless, it is important to clarify velocity dispersion created by (non-collective)
two-body relaxation in a disc potential, 
because the two-body relaxation is one of the most basic processes 
in a self-gravitating disc system.
Even if the collective effects dominate the disc heating in our galaxy, 
it is important to understand a competitive process, the heating by two-body 
relaxation.
Furthermore, the two-body relaxation may domminate in the central region 
of our galaxy or in other galaxies.
In the present paper, we are concerned with dynamics regulated 
by the two-body relaxation in an 'idealized' disc system 
with the potential characterized by a parameter $\kappa/\Omega$, 
where $\kappa$ and $\Omega$ are horizontal epicycle frequency (see Eqs.\ (\ref{epifre})) 
and circular frequency of the disc potential, 
which indicates radial dependence of a disc potential.

\begin{figure}
 \epsfbox{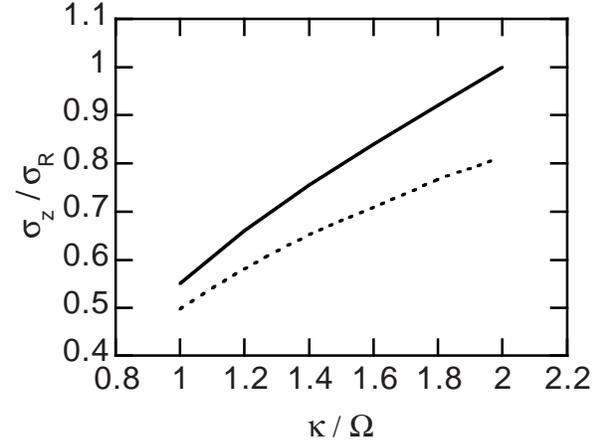}
 \caption{The equilibrium ratio $\sigma_z / \sigma_R$ as a function of $\kappa/\Omega$ 
obtained by Lacey(1984) (solid line) and IKM93 (dotted line).
Since the result of IKM93 is weakly dependent on the disc scale height owing 
to their choice of $\ln\Lambda$, 
we plot the case in which  the disc scale height is larger than the 
tidal radius of a particle where their assumption is valid
(the case where $\irmsj =5$, see section~2).}
 \label{fig1}
\end{figure}

It is suggested that the equilibrium ratio ($\sigma_{z}/\sigma_{R}$)
depends on $\kappa/\Omega$.
Lacey(1984) and Ida, Kokubo \& Makino(1993)
(hereafter IKM93) analytically investigated the equilibrium value of 
$\sigma_{z}/\sigma_{R}$ as a function of $\kappa/\Omega$. 
They assumed that the evolution of the velocity dispersion is described 
by the sum of many independent two-body scatterings in a disc potential
with various initial conditions.
They adopted the epicycle approximation (e.g. Binney \& Tremaine 1987)
and calculated orbital changes with the impulse approximation 
neglecting the external disc potential. 
IKM93 also suggested that the anisotropy is produced by 
deceleration of horizontal velocity at close approach due to shear motion
(see section~2.2).
For the Keplerian potential ($\kappa/\Omega = 1$) and the 
galactic potential in the solar neighbourhood($\kappa/\Omega \simeq 1.4$), 
IKM93 obtained the value of 
$\sigma_{z}/\sigma_{R}$ that are consistent with  
$N$-body simulations (and observations), while Lacey(1984) show significantly large values
(Fig.\ \ref{fig1}).
IKM93 claimed that choice of the maximum impact parameter in the two-body 
formulae affects the equilibrium value of $\sigma_{z}/\sigma_{R}$.
IKM93 carefully chose the maximum impact parameter comparing with numerical 
orbital integration to obtain smaller $\sigma_{z}/\sigma_{R}$ than that of Lacey(1984).
However, IKM93 obtained $\sigma_{z}/\sigma_{R} \neq 1$  in the 
limit of $\kappa/\Omega = 2$ (solid-body rotation).
It would be reasonable to consider $\sigma_{z}/\sigma_{R} = 1$ 
in the solid-body rotation, since shear motion vanishes. 
Lacey(1984) gave an argument with Jeans theorem to show 
$\sigma_{z}/\sigma_{R} = 1$ in that case.
Lacey(1984) obtained $\sigma_{z}/\sigma_{R} = 1$ for $\kappa/\Omega = 2$, 
which is consistent with the above argument, 
although he failed to reproduce consistent values of $\sigma_{z}/\sigma_{R}$
for smaller $\kappa/\Omega$.
Things have been obscured because of the lack of both observation and 
numerical work in the limiting case, $\kappa/\Omega \simeq 2$.

To address this problem, we performed numerical simulations in 
disc potentials with wide range of $\kappa/\Omega$ up to $\sim 2$.
In section~2, we calculate the disc heating as the sum of 
many independent two-body 
scatterings in a disc potential, as Lacey(1984) and IKM93 did, but 
orbital changes are obtained by numerical orbital integrations.
Our numerical calculations show an isotropic-dispersion ($\sigma_{z}/\sigma_{R} \sim 1$)
regime in velocity space if $\kappa/\Omega \ga 1.5$.
We also found that the range of the regime expands
with $\kappa/\Omega$ and the regime dominates all over the velocity 
space in the limiting case of the solid-body rotation 
($\kappa/\Omega = 2$).

In section~3, to confirm 
our results  obtained in section~2, we performed $N$-body simulations of particles 
in various disc potentials.
The results in section~2 and 3 are in good agreement with each other.
In section~4, we summerize our results.

\section{VELOCITY EVOLUTION DUE TO MANY TWO-BODY SCATTERINGS 
IN DISC POTENTIALS}

\subsection{Basic Formulation}

In this section, we consider a swarm of test bodies (particle~1) gravitationally perturbed by 
field bodies (particle~2) in a disc potential field.
We evaluate change rate of the velocity dispersion of the test bodies 
by statistically compiling the velocity changes in 
individual two-body encounters with field bodies, 
which are calculated numerically,  
following the method adopted by Ida (1990) and Kokubo \& Ida (1992).
Here we are also concerned with the cases with $\kappa/\Omega$ close to 2
while Ida (1990) and Kokubo \& Ida (1992) only studied the cases with $\kappa/\Omega = 1.0$ and 
1.39.

We assume that 'background bodies' generating the disc potential 
is continuously distributed and they do not contribute to 
two-body scattering.
   
We adopt the epicycle approximation 
(see e.g. Petit \& H\'{e}non 1986; Binney \& Tremaine 1987):  
the velocity dispersion is sufficiently smaller than rotational 
velocity around galactic centre.
We use the following rotating coordinates:
\eq
	\left\{
\begin{array}{lc}
x = &
\disp 
(R - a)/r_{\rm g}, \\
y = &
\disp 
(a\phi - a\Omega t)/r_{\rm g},\\
z = &
\disp 
z/r_{\rm g},
\end{array}
        \right.
\eqend
where $(R,\phi,z)$ are cylindrical coordinates centred at the bottom 
of the disc potential, 
$\Omega$ is the circular frequency at $R = a$.
The normalization factor $r_{\rm g}$ is defined by
\eq
r_{\rm g} = \left(\frac{G(m_1 + m_2)}{\Omega^2}\right) 
^{\frac{1}{3}},
\label{hillradius}
\eqend
where $m_1$ and $m_2$ are the masses of the particle~1 and 2 and 
$G$ is the gravitational constant.
The radius $r_{\rm g}$ corresponds to tidal radius within a numerical factor of 
$O(1)$ except for $\kappa/\Omega \simeq 2$ (see Eq. (\ref{tidalradi}))

In the epicycle approximation, the unperturbed orbits are given by 
(Petit \& H\'{e}non 1986; Binney \& Tremaine 1987)
\eq
\left\{ \begin{array}{l}
\disp
x_j = b_j - e_j \frac{\Omega}{\kappa} 
\disp
\cos (\frac{\kappa}{\Omega} t - \tau_j), \\
\disp
y_j = \lambda_j - \frac{\alpha}{2}b_j t  + 
\disp
2 e_j \frac{\Omega^2}{\kappa^2} \sin (\frac{\kappa}{\Omega} t - \tau_j), \\
\label{eq2120}
\disp
z_j = i_j \frac{\Omega}{\nu} \sin(\frac{\nu}{\Omega} t - \omega_j),
\end{array} \right.  
\eqend
and 
\eq
\left\{ \begin{array}{l}
\disp
\dot{x_j} = e_j \sin (\frac{\kappa}{\Omega} t - \tau_j), \\
\disp
\dot{y_j} = - \frac{\alpha}{2}b_j + 
\disp
2 e_j \frac{\Omega}{\kappa} \cos (\frac{\kappa}{\Omega} t - \tau_j), \\
\label{eq2121}
\disp
\dot{z_j} = i_j \cos(\frac{\nu}{\Omega} t - \omega_j),
\end{array} \right.  
\eqend
where time is scaled by $\Omega^{-1}$, and 
$\kappa$ and $\nu$ are epicycle frequency and frequency of vertical oscillation
 which are defined by
\eq 
\left\{
\begin{array}{l}
\disp
\kappa = \left(\frac{\partial^2 \Phi}{\partial R^2}\right)_{R=a} + 
\disp
3 \Omega^2 = 
\disp
2\Omega \left(R \frac{\partial \Omega}{\partial R}\right)_{R=a},\\
\disp
\label{epifre}
\nu = \left(\frac{\partial^2 \Phi}{\partial z^2}\right)_{R=a},
\end{array}
\right.
\eqend
where $\Phi(R,z)$ is an axisymmetric disc potential.
For convenience, we introduced a parameter $\alpha$ which indicates the strength of shear motion:
\eq 
\disp
\alpha = 4 - \frac{\kappa^2}{\Omega^2}. 
\label{alpha}
\eqend
The quantities 
$e$, $i$, $b$, $\tau$, $\omega$, and $\lambda$ are the constants 
of integration.
Equations (\ref{eq2120}) and (\ref{eq2121}) represent the particle 
motion as a combination of planar epicycle and vertical oscillation
around the guiding centre rotating in a non-inclined circular orbit. 
The quantities $e \Omega /\kappa$
 and $\tau$ are the amplitude and the phase of the horizontal 
oscillation, respectively. 
Similarly, $i \Omega /\nu$ and $\omega$ are those  
of the vertical oscillation.
The first term of the right hand side of $\dot{y_j}$ represents the  shear velocity, 
$[\Omega(a + r_{\rm g} b) - \Omega(a)]a$ scaled by  $r_{\rm g}\Omega$ 
(we assumed $b r_{\rm g} \ll a$).
In the special case of Kepler rotation ($\kappa/\Omega = 1.0$), 
the constants $e r_{\rm g} /a$ and $i r_{\rm g} /a$ are called eccentricity and inclination.

From equations (\ref{eq2121}),
rms velocity averaged over an epicycle period of $j$-th 
particle is given by 
\eq
\left\{
\begin{array}{l}
\disp
\sigma_{jR} \equiv \bar{\dot{x_j^2}}^{\frac{1}{2}} r_{\rm g} \Omega= 
\disp
\frac{e_j}{\sqrt{2}} r_{\rm g} \Omega,\\
\disp
\sigma_{jz} \equiv \bar{\dot{z_j^2}}^{\frac{1}{2}} r_{\rm g} \Omega= 
\disp
\frac{i_j}{\sqrt{2}} r_{\rm g} \Omega,
\end{array}
\right.
\eqend
where '$\;\bar{\;}\;$' denotes time-averaging during the epicycle period.  
Then we define the velocity dispersion of a swarm of particle~1 (test bodies) as 
\eq
\left\{
\begin{array}{l}
\disp
\sigma_{R} \equiv 
\disp
\frac{\langle e_1^2 \rangle^{1/2}}{\sqrt{2}} r_{\rm g} \Omega, \\
\label{defsigma}
\sigma_{z} \equiv 
\disp
\frac{\langle i_1^2 \rangle^{1/2}}{\sqrt{2}} r_{\rm g} \Omega, \\
\end{array}
\right.
\eqend
where '$\langle \;\; \rangle$' denotes ensemble-averaging
 (i.e. $\sigma_{R}= \langle \sigma_{1R}^2 \rangle^{1/2}$ and 
$\sigma_{z}=\langle \sigma_{1z}^2 \rangle^{1/2}$)
We also call $\langle e_1^2 \rangle^{1/2}$ and $\langle i_1^2 \rangle^{1/2}$ 
'normalized velocity dispersion' and pursue evolution of them.
Kokubo \& Ida (1992) considered gravitational scatterings of 
test bodies (disc stars) by  many massive field bodies 
(giant molecular clouds) which are in non-inclined circular orbit.
According to them, the evolution of the normalized velocity dispersion
of the test bodies are given by 
\begin{equation}
        \left\{
\begin{array}{l}
\disp
\frac{{\rm d}\langle e_1^2 \rangle}{{\rm d}t} = 
\disp
n_{s2} r_{\rm g}^2 \Omega \int f_1(e_1,i_1)P_{\rm heat}(e_1,i_1)\,{\rm d}e_1^2 \,{\rm d}i_1^2, \\
\label{evolution5} 
\disp
\frac{{\rm d}\langle i_1^2 \rangle}{{\rm d}t} = 
\disp
n_{s2} r_{\rm g}^2 \Omega \int f_1(e_1,i_1)Q_{\rm heat}(e_1,i_1)\,{\rm d}e_1^2 \,{\rm d}i_1^2   
\end{array}
\right.
\end{equation}
where 
\eq
\left\{
\begin{array}{l}
\displaystyle  
P_{\rm heat}(e,i) \equiv 
\disp
\int \Delta e^2 \frac{\alpha}{2} |b|
\disp
\,\frac{{\rm d}\tau {\rm d}\omega}{(2\pi)^2}\,{\rm d}b = \int p(e,i,b){\rm d}b, \\
\label{hrate}
\disp
Q_{\rm heat}(e,i) \equiv
\disp
 \int \Delta i^2 \frac{\alpha}{2} |b|
\disp
\, \frac{{\rm d}\tau {\rm d}\omega}{(2\pi)^2}\,{\rm d}b = \int q(e,i,b){\rm d}b
\end{array}
\right.
\eqend
(we assumed $b$, $\tau$, and $\omega$ are distributed randomly).
The quantities $p(e,i,b)$ and $q(e,i,b)$ are introduced for later convenience and written as 
\eq 
\left\{
\begin{array}{l}
\disp
p(e,i,b) \equiv 
\disp
\int \Delta e^2 \frac{\alpha}{2} |b| \frac{{\rm d}\tau {\rm d}\omega}{(2\pi)^2}, \\ 
\disp
\label{smallpq}
q(e,i,b) \equiv 
\disp
\int \Delta i^2 \frac{\alpha}{2} |b| \frac{{\rm d}\tau {\rm d}\omega}{(2\pi)^2}.
\end{array}
\right.
\eqend
In Eqs.\ (\ref{evolution5}), $n_{s2}$ is the surface number density of 
the massive bodies and $f_1$ is the velocity distribution of 
the test bodies, which is normalized as 
\eq 
\int f_1(e,i) {\rm d}e^2 {\rm d}i^2 = 1.
\eqend
Equations (\ref{evolution5}) are valid when $m_2 \gg m_1$.
In more general case where  
the mass of a test body (particle~1)
is comparable to that of a field body (particle~2),
Eqs. (\ref{evolution5}) are revised as
\begin{equation}
        \left\{
\begin{array}{l}
\disp
\frac{{\rm d}\langle e_1^2 \rangle}{{\rm d}t} = 
\disp
n_{s 2} r_{\rm g}^2 \Omega \left(\frac{m_2}{m_1 + m_2}\right)^2 
\disp
\langle P_{\rm heat}\rangle, \\
\label{evolution6} 
\disp
\frac{{\rm d}\langle i_1^2 \rangle}{{\rm d}t} = 
\disp
n_{s 2} r_{\rm g}^2 \Omega \left(\frac{m_2}{m_1 + m_2}\right)^2 
\disp
\langle Q_{\rm heat}\rangle,
\end{array}
\right.
\end{equation} 
where 
\eq
\left\{
\begin{array}{l}
\disp
\langle P_{\rm heat}\rangle = \int f(e,i)P_{\rm heat}(e,i){\rm d}e^2 {\rm d}i^2, \\
\label{heatave}
\disp    
\langle Q_{\rm heat}\rangle = \int f(e,i)Q_{\rm heat}(e,i){\rm d}e^2 {\rm d}i^2 \\ 
\end{array}
\right.
\eqend
(Ohtsuki 1998, Stewart \& Ida 1998).
In Eqs.\ (\ref{heatave}), $e$ and $i$ are velocity dispersion of the 
relative motion defined by
\eq
\left\{
\begin{array}{l}
e \cos \tau = e_2 \cos \tau_2 - e_1 \cos \tau_1, \\
e \sin \tau = e_2 \sin \tau_2 - e_1 \sin \tau_1, \\
i \cos \omega = i_2 \cos \omega_2 - i_1 \cos \omega_1, \\
i \sin \omega = i_2 \sin \omega_2 - i_1 \sin \omega_1. 
\end{array}
\right.
\eqend
Then, we obtain
\eq
\left\{
\begin{array}{l}
\langle e^2 \rangle = \langle e_1^2 \rangle + \langle e_2^2 \rangle, \\
\label{erelation}
\langle i^2 \rangle = \langle i_1^2 \rangle + \langle i_2^2 \rangle.
\end{array}
\right.
\eqend  
Under an assumption that both $f_1$ and $f_2$ 
(velocity distributions of particles~1 and 2) are Rayleigh distribution, 
$f$ is again Rayleigh distribution (Ohtsuki 1998, Stewart \& Ida 1998):
\eq
\disp
\begin{array}{ll}
f(e,i){\rm d}e^2 {\rm d}i^2 &= 
\disp
\frac{1}{\eemean \iimean} 
\label{Rayleigh}
\disp
\exp\left(-\frac{e^2}{\eemean} -\frac{i^2}{\iimean}\right) {\rm d}e^2 {\rm d}i^2 \\
&= 
\disp
\frac{4ei}{\eemean \iimean} 
\disp
\exp\left(-\frac{e^2}{\eemean} -\frac{i^2}{\iimean}\right) {\rm d}e {\rm d}i.
\end{array}
\eqend 
Equations (\ref{evolution6}) and (\ref{heatave}) imply that   
$\langle P_{\rm heat} \rangle$ and $\langle Q_{\rm heat} \rangle$,
which are calculated only by orbital change in the relative motion,
determine the ratio of the velocity dispersion.
It should be noted that integration with $i$ in Eqs. (\ref{heatave})
is also equivalent to that with $v_z$ and $z$, i.e., averaging with 
vertical velocity and height from 
disk mid-plane (Lissauer \& Stewart 1993).   
Strictly speaking, Eq.\ (\ref{evolution6}) should include the terms 
expressing recoil of dynamical friction, which is proportional to
$(m_1 \langle e_1^2 \rangle - m_2\langle e_2^2 \rangle$) or
$(m_1 \langle i_1^2 \rangle - m_2\langle i_2^2 \rangle$)
(Ohtsuki 1998; Stewart \& Ida 1998), 
however, we neglect it, assuming the energy equipartition between
particles~1 and 2 is already realized.
In this case, it is shown that the recoil terms are much smaller than the heating terms 
(right hand sides of Eqs. (13)) if $m_1 \ll m_2$ (Ida 1990).
Furthermore, in the identical particle case, Eq.(\ref{evolution6}) is exact. 

The equations of the relative motion are given by 
(e.g. Icke 1982; Petit \& H\'{e}non 1986; Kokubo \& Ida 1992)  
\eq
\left\{ \begin{array}{llll}
\ddot{x} - 2 \dot{y} & = & 
\alpha x &
\disp
- x/r^3 ,\\
\ddot{y} + 2 \dot{x} & = & &
\disp 
- y/r^3 ,\\
\label{epieqrel}
\disp 
\ddot{z}             & = & 
-(\nu/\Omega)^2 z &
\disp
- z / r^3,
\end{array}
\right.
\eqend 
where time is scaled by $\Omega^{-1}$,
$(x,y,z) = (x_2 -x_1,y_2-y_1,z_2 - z_1)$, and 
$r$ is scaled distance, $r=(x^2+y^2+z^2)^{1/2}$.
The last terms on the right hand side of Eqs.\ (\ref{epieqrel}) 
represent the gravitational interaction between two particles.
Since we scale length and time by $r_{\rm g}$ and $\Omega$, $G(m_1+m_2)x/r^3$ is 
reduced to $x/r^3$ in these equations.
The terms $-2\dot{y}$ and $ 2\dot{x}$ are the Coriolis force. 
In the case of $\kappa/\Omega = 1.0$ and $\nu/\Omega = 1.0$, 
Eqs.\ (\ref{epieqrel}) are called Hill's equations 
which describe motion in the Kepler potential.

\begin{figure}
\epsfbox{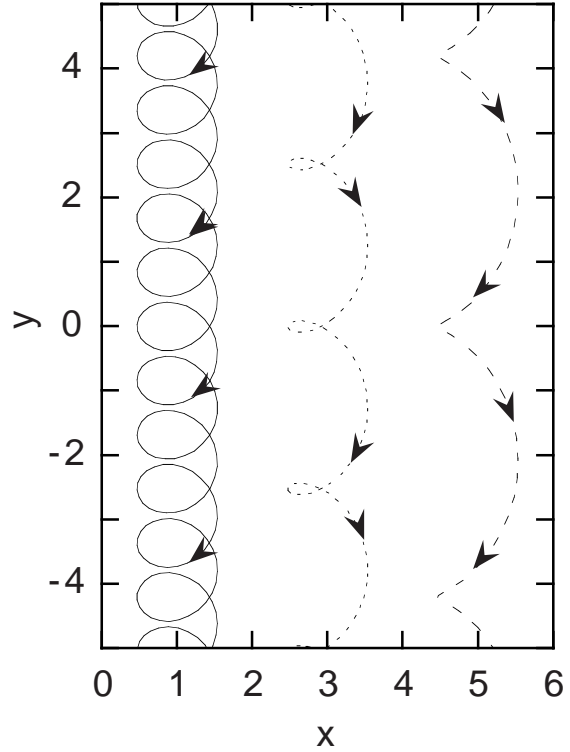}	
 \caption{Examples of the unperturbed orbit on the $x$-$y$ plane. 
For all cases, $e = 1.0$ and $\kappa/\Omega = 1.87$.
Solid, dotted, and dashed lines are orbits with $b=1.0$, 
$b=3.0$, and $b=5.0$, respectively.}
 \label{fig2}
\end{figure}

We will numerically integrate Eqs. (\ref{epieqrel}) to evaluate 
the changes of orbital elements, in particular, $e$ and $i$.
The relative orbital elements are related to relative motion 
$(x,y,z,\dot{x},\dot{y},\dot{z})$
in a similar way to Eqs. (\ref{eq2120}) and (\ref{eq2121}) as  
\eq
\left\{ \begin{array}{l}
\disp 
x = b - e \frac{\Omega}{\kappa} \cos (\frac{\kappa}{\Omega} t - \tau),\\
\disp 
y = \lambda - \frac{\alpha}{2}b t + 
\disp 
    2e \frac{\Omega^2}{\kappa^2} \sin (\frac{\kappa}{\Omega} t - \tau), \\
\label{xrelkai}
\disp 
z = i\frac{\Omega}{\nu} \sin(\frac{\nu}{\Omega} t - \omega),
\end{array} \right.
\eqend
and
\eq
\left\{ \begin{array}{l}
\disp 
\dot{x} = e \sin (\frac{\kappa}{\Omega}t - \tau), \\
\disp 
\dot{y} = - \frac{\alpha}{2}b + 
\disp 
   2e \frac{\Omega}{\kappa} \cos (\frac{\kappa}{\Omega} t - \tau), \\
\label{vrelkai}
\disp    
\dot{z} = i \cos(\frac{\nu}{\Omega} t - \omega).
\end{array} \right.
\eqend
When the two particles are so far away that the mutual gravitational
terms in Eqs.\ (\ref{epieqrel}) are negligible, the relative orbital 
elements are constants.
When they approach each other, the orbital elements change through the 
mutual perturbation.
We show the examples of unperturbed motion, 
i.e., the motion with constant orbital elements in Fig.\ \ref{fig2}. 
In this figure, $\kappa/\Omega = 1.87$, $e = 1.0$, and $b =  1.0$ (solid line), 
3.0 (dotted line), and 5.0 (dashed line). 

\subsection{Numerical Results of $P_{\rm heat}$ and $Q_{\rm heat}$}

We first investigate the behaviour of the 'elementary' quantities 
$P_{\rm heat}$ and $Q_{\rm heat}$ (Eqs. (\ref{hrate})) 
since they 
show clearer physical properties than $\langle P_{\rm heat} \rangle$ and
 $\langle Q_{\rm heat} \rangle$
and the averaging 
(Eqs. (\ref{heatave})) will not change the ratio of 
the velocity dispersion substantially.

To obtain $P_{\rm heat}$ ($Q_{\rm heat}$), we calculate $\Delta e^2$ 
($\Delta i^2$) with various sets of $(b,\tau,\omega)$ 
through orbital integration for each set of $(e,i)$, 
following Ida (1990) and Kokubo \& Ida (1992).
As described in the previous subsection, only relative motion which obeys 
Eqs. (\ref{epieqrel}) is calculated. 

When relative distance $r$ is large enough 
that mutual gravitation can be neglected, the orbital elements 
$(e,i,b,\tau,\omega,\lambda)$ are constants. 
We start our orbital integration with sufficiently large $y$.
A body approaches the other owing to shear motion.
During the encounter, the orbital elements are changed by mutual gravitational 
perturbation.
We stop the integration when $|y|$ becomes large enough again.
Changes of the orbital elements are determined as the difference between 
orbital elements of initial and final states, i.e., 
$\Delta e^2 = e_{\rm final}^2 - e_{\rm initial}^2$.
Since contribution in the integral (\ref{hrate}) from non-crossing distant 
encounters rapidly decreases with $|b|$ in a disc system 
(Ida 1990; Hasegawa \& Nakazawa 1990; Kokubo \& Ida 1992), initial $b$ of orbits we calculated 
is confined in some finite regime. 
Furthermore, according to the symmetry of basic equations, 
the changes of $e^2$ and $i^2$ take the same values
for the orbits with $b$ and $-b$ and those with $\omega$ and $\omega + \pi$.
Hence, we calculated orbits with $0 < b \leq b_{\rm max}$, $0 \leq \omega \leq \pi$,
and $-\pi \leq \tau \leq \pi$ (for the value of $b_{\rm max}$, see below).

The orbits are integrated with the fourth-order Hermite scheme
(Makino \& Aarseth 1992).
We also employed the algorithm
developed by Emori, Ida, \& Nakazawa (1993), where 
the part of deviation from the unperturbed epicycle orbit is numerically
calculated while the part of the unperturbed epicycle orbit is analytically
calculated.

We numerically calculated the heating rates with various $\kappa/\Omega$:
$\kappa/\Omega = 1.00$, $1.30$, $1.58$, $1.73$, and $1.87$.
Since IKM93 showed that anisotropy in the velocity dispersion 
is closely related to  
the shear motion, which depends on $\kappa/\Omega$ but not on $\nu/\Omega$,
the parameter $\nu/\Omega$ is fixed to $1.0$.
In the limit of $\kappa/\Omega \rightarrow 2.0$, shear motion vanishes 
so that orbital integration becomes difficult. 
The case with $\kappa/\Omega \sim 1$ is well investigated by 
several authors (e.g. Ida 1990, Kokubo \& Ida 1992), hence
we are mainly concerned with  
the results with $\kappa/\Omega = 1.58$, $1.73$, and $1.87$.
 
\begin{figure*}
  \epsfbox{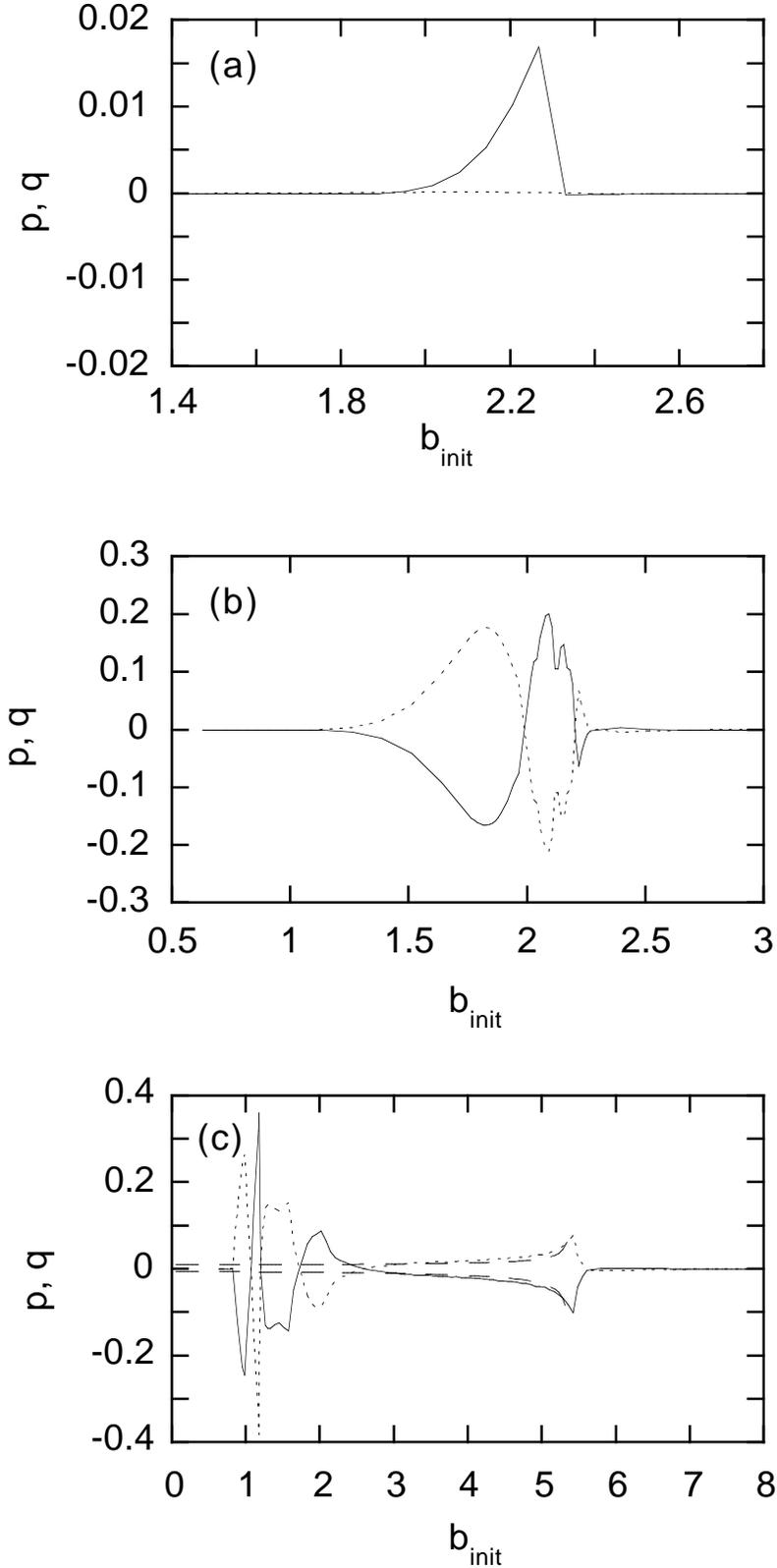}
  \caption{Dependence of $p(e,i,b)$ (solid lines) and 
$q(e,i,b)$ (dotted lines) on initial $b$. 
(a) Case with $(e,i) = (0.19,0.19)$.
The $b$-dependence is plotted every $0.05$ ($=\delta \,b$).
To evaluate one point, $50 \times 25$ orbits with different $\tau$ and $\omega$ are 
integrated. 
(b) Case with $(e,i) = (1.23,1.23)$. 
In this case, we varied $\delta b$ according to $b$-dependence of $p$ or $q$
(we took smaller $\delta b$ where $p$ or $q$ rapidly changes):
$\delta b = 0.01 \mbox{-} 0.1$. 
(c) Case with $(e,i) = (10.1,5.04)$. 
As well as the case (b), we varied $\delta b$ according to 
$b$-dependence of $p$ or $q$ so that $\delta b = 0.01 \mbox{-} 0.2$.
For each $b$, we calculated $100 \times 50$ - $400 \times 200$ orbits 
with different $\tau$ and $\omega$. 
We also plot IKM93's result (dashed lines).}
\label{fig3}
\end{figure*}

First we show the results for $\kappa/\Omega = 1.87$.
In this case, we calculated $P_{\rm heat}$ and $Q_{\rm heat}$ for 392 sets of 
$e$ and $i$.
For each set of $(e,i)$, we integrated $10^4$ - $10^6$ orbits
with different sets of $(b,\tau,\omega)$.
In Figs.\ \ref{fig3}-a to \ref{fig3}-c, we show obtained
$p(e,i,b)$ (solid lines) and $q(e,i,b)$ (dotted lines) as functions of $b$. 
Integration of $p$ and $q$ with $b$ yields $P_{\rm heat}$ and $Q_{\rm heat}$
(Eqs. (\ref{hrate})).
In Figs.\ \ref{fig3}-a, b, and c, 
($e,i$) = (0.19,0.19), (1.23,1.23), and (10.1,5.04), 
respectively.
Individual figures correspond to the results in the shear dominant
region, the horseshoe dominant region, and the dispersion dominant region, 
which we define below.
These figures suggest that only intermediate $b$ 
contributes to the heating. 
The orbits with the intermediate $b$ can closely approach each other.
The orbits with smaller $b$ turns back at distant $y$ by the Coriolis 
force ('horseshoe orbits' (Brown 1911)), 
while those with larger $b$ pass by at distant $x$.
Ida (1990) and Kokubo \& Ida (1992) showed that cumulative contribution 
($P_{\rm heat}$ and $Q_{\rm heat}$)
of distant encounters with large $b$ is negligible even if it is integrated 
over $b$ to infinity.  

In the case of large $e$, the range of the 
strongly perturbed orbits in the $b$-space is extended 
by large amplitude of radial excursion, $e\Omega/\kappa$, as shown 
in Fig.\ \ref{fig3}.
In the unperturbed orbits, the condition for orbit crossing is
$\mid b \mid < e\Omega/\kappa$ (Eq.(\ref{xrelkai})).
Hence we usually calculate orbits in the range of 
$0.6 \la b \la e\Omega/\kappa + 2.5$.
In the cases of Figs.\ \ref{fig3}-a, b, and c, 
we calculated $3 \times 10^4$, $7 \times 10^4$, and 
$4.3 \times 10^6$ orbits. 
The numbers of calculated orbits are large enough that
the integrated values of $P_{\rm heat}(e,i)$ and $Q_{\rm heat}(e,i)$
change by less than $5$-$10$ per cent by the choice of calculated
sets of initial conditions.

\begin{figure*}
  \epsfbox{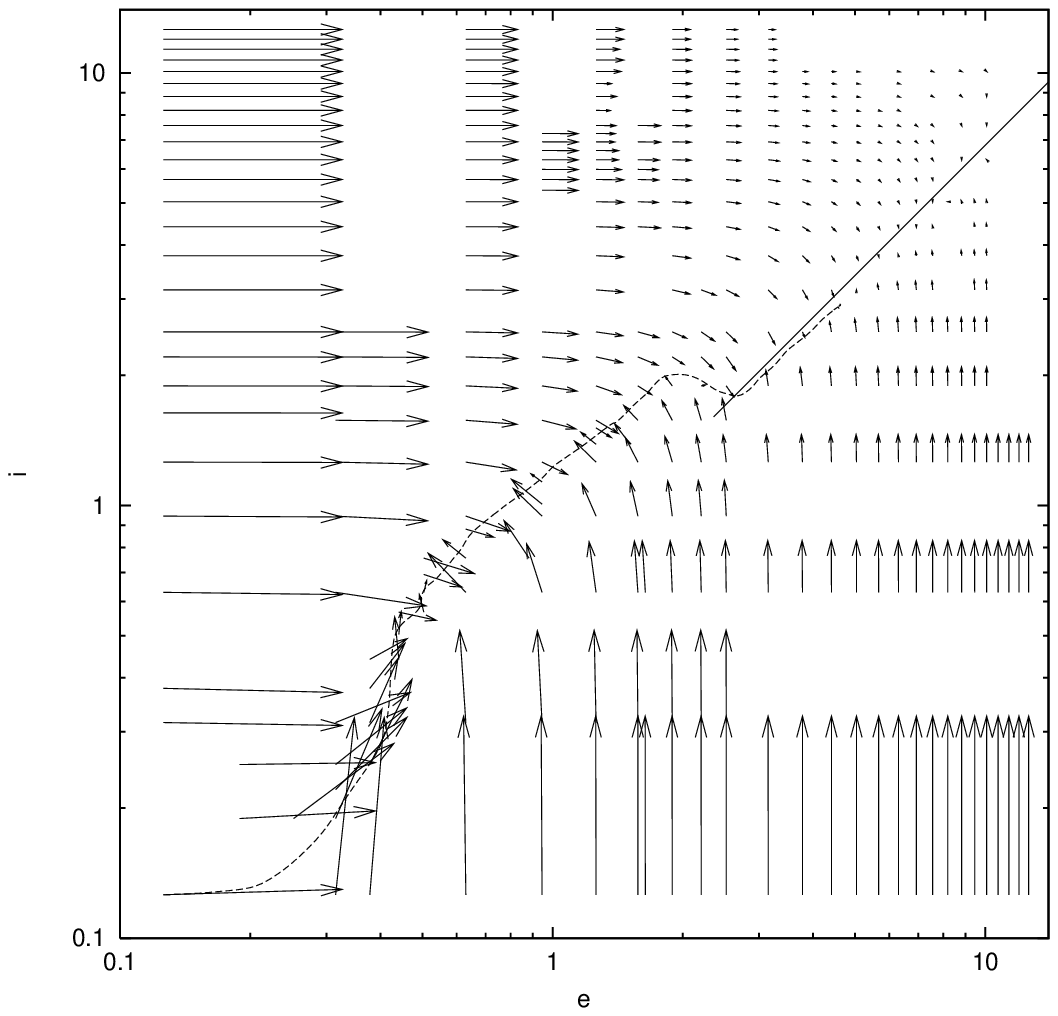}
  \caption{Directions and magnitude of the evolution of $e$ and $i$
on the $e\mbox{-}i$ plane in the case where $\kappa/\Omega = 1.87$.
The angles of the vectors are given by $\tan^{-1}( Q_{\rm heat}e^2 / P_{\rm heat}i^2)$
 and the length is $\mbox{max}(a, a\log[b(|P|/e^2 + |Q|/i^2 + 1)])$, 
where $a = 0.2$ and $b = 200$.
Dashed line denotes the trajectory with initial condition 
$(e,i) = (0.126,0.126)$.
Solid line denotes the line $i = 0.68e$ which is predicted by IKM93.}
\label{fig4}
\end{figure*}

In Fig.\ \ref{fig4}, we compile the calculated results for 392 sets of $(e,i)$ with vectors.
The angle between a vector and $e$-axis is determined by 
\[
\theta = \tan^{-1} \left( \frac{Q_{\rm heat}/i^2}{P_{\rm heat}/e^2} \right),
\]
while the length of the vectors are defined as 
$a \log[b(|P|/e^2 + |Q|/i^2 + 1)]$, 
where $a$ and $b$ are constants chosen for the vectors
to be easily seen (in Fig.\ \ref{fig4}, $a = 0.2$ and $b=200$),
and the factor 1 is introduced so that the length is always positive.
These vectors show evolution trend of $e$ and $i$.
If a vector points to upper-right direction ($\theta = 45^{\circ}$), 
$e$ and $i$ increase keeping $i/e$ constant.
Since 
\[
\disp
\frac{{\rm d}(i/e)}{{\rm d}t} = \frac{i}{2e}(\frac{1}{i^2}\frac{{\rm d}i^2}{{\rm d}t} 
\disp
- \frac{1}{e^2}\frac{{\rm d}e^2}{{\rm d}t}),
\]
$i/e$ is kept constant when 
\[
\frac{1}{i^2}\frac{{\rm d}i^2}{{\rm d}t} = \frac{1}{e^2}\frac{{\rm d}e^2}{{\rm d}t}
\]
or equivalently, 
\[ 
Q_{\rm heat}/i^2 = P_{\rm heat}/e^2.
\]
The evolution of $e$ and $i$ is divided into two steps.
The first evolution is relatively rapid evolution toward the equilibrium state of 
$i/e$, and the second one is the gradual increase in $e^2 + i^2$ 
keeping $i/e$ constant.
Fig.\ \ref{fig4} suggests that the first step is faster than the second one.
In the case where $e$ and $i$ are sufficiently large, we can quantitatively show the 
time-scale of the first step is much shorter than that of the second one (Appendix A).
As a result, $e$ and $i$ would evolve along the line where vectors gather.
Since ${\rm d}e^2/{\rm d}t = C P_{\rm heat}$ and 
${\rm d}i^2/{\rm d}t = C Q_{\rm heat}$, where $C$ is some constant, 
we can integrate typical trajectory of a particle on the 
$e$-$i$ plane using the data in Fig.\ \ref{fig4} and their interpolation.
The trajectory with initial condition $(e,i) = (0.126,0.126)$
is plotted as dashed line in Fig.\ \ref{fig4}.
Since the trajectory reaches the equilibrium state of $i/e$ rapidly, it  
corresponds to a set of $(e,i)$ in the 'equilibrium' state. 

Fig.\ \ref{fig4} shows three different regions according 
to the manner of the evolution of 
$e$ and $i$, in particular, equilibrium ratio $i/e$: 
{\em shear dominant region} $(e,i \la 0.4)$, 
{\em horseshoe dominant region} $(0.4 \la e,i \la 2)$, and 
{\em dispersion dominant region} $(e,i \ga 2)$.
In each region, the equilibrium ratio $i/e$ is $\ll 1$, $\sim 1.1$, and $\sim 0.7$, 
respectively, in the case where $\kappa/\Omega=1.87$.

\begin{figure*}
 \epsfbox{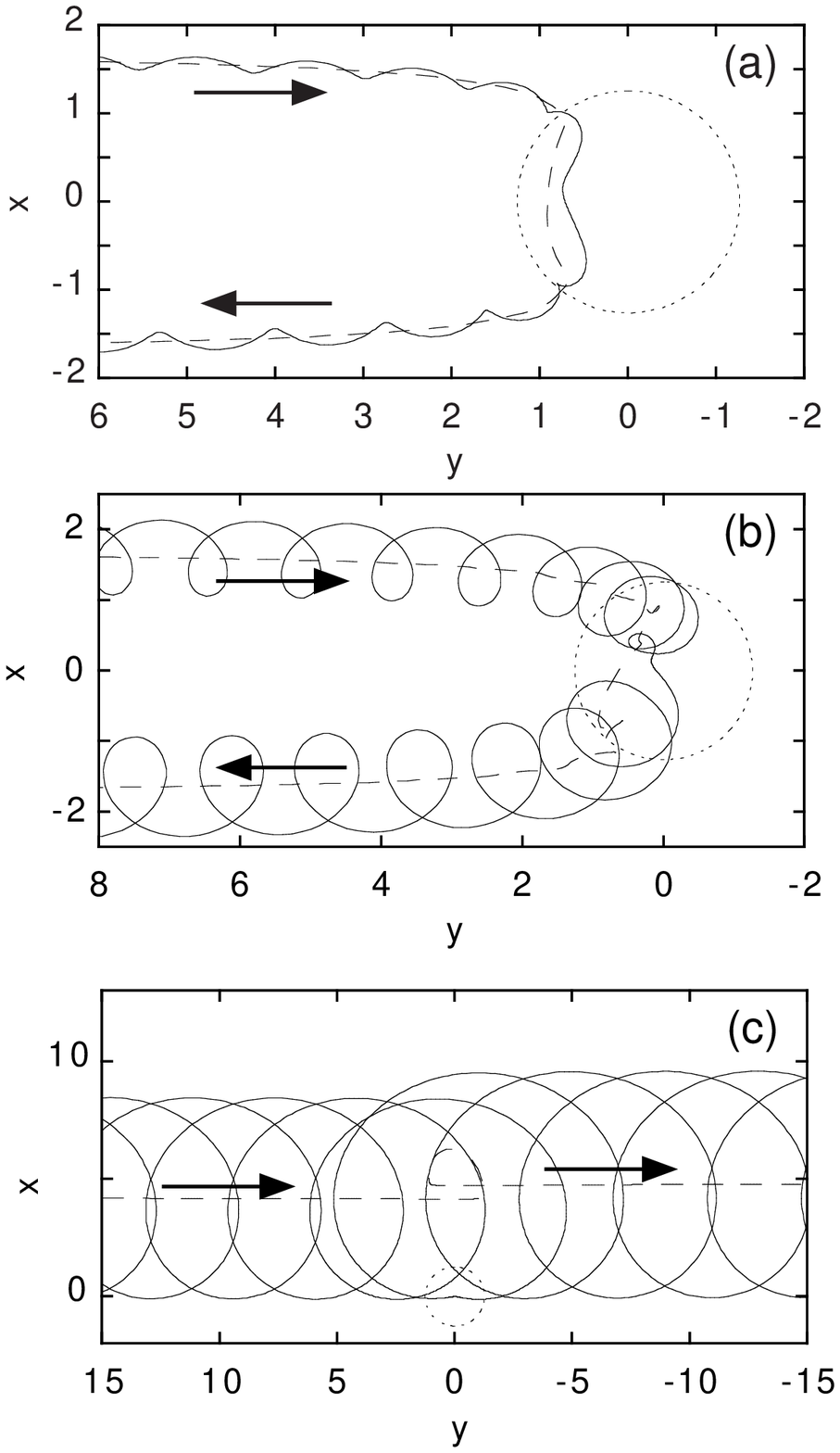}
 \caption{Examples of orbit in the shear dominant region (a), 
the horseshoe dominant region (b), and the dispersion dominant region (c)
in the case where $\kappa/\Omega = 1.87$.
The solid lines denote the orbit, the dashed lines the guiding-center motion, 
and the circle with dotted line represents the tidal sphere ($r = r_{\rm t}$).
Arrows represents the directions of the guiding-center motion.
In each figure, initial orbital elements are 
$(e,i,b) = (0.19,0.19,2.14)$ in (a), $(e,i,b) = (1.23,1.23,2.14)$ in (b), 
and $(e,i,b) = (10.1,5.04,4.2)$ in (c) (we omitted the other orbital elements 
such as phase variables).}
 \label{fig5}
\end{figure*}

\vspace{0.5cm}
{\em shear dominant region} 
\vspace{0.5cm}

In the shear dominant region, the vertical and horizontal 
epicycle amplitude are small
and shear motion dominates. 
Since shear motion is horizontal and orbits are bent before
the orbits come close to each other,
gravitational scattering takes place two-dimensionally. 
Accordingly, $i$ hardly changes as shown
in Fig.\ \ref{fig3}-a.
Dominant perturbation to $e$ comes from 
orbits with $b \sim 2$.
These orbits are distant 'horseshoe'-type encounters 
as shown in Fig.\ \ref{fig5}-a.

The orbital behaviour changes where  
the epicyclic oscillation velocity, $(e^2 + i^2)^{1/2} \sim \sqrt{2}\, e$,
is comparable to the shear velocity, 
namely, 
where $\alpha b /2   \sim \sqrt{2}\, e$ (see Eqs.\ (\ref{vrelkai})). 
Hence the boundary should be $e \sim 0.35$, which is consistent with Fig.\ \ref{fig4}.

\vspace{0.5cm}
{\em dispersion dominant region} 
\vspace{0.5cm}

When $e,i \ga \alpha$, approach velocity is dominated by the random velocity 
$v = (e^2 + i^2)^{1/2} r_{\rm g}\Omega$ 
rather than the shear velocity ($\alpha b r_{\rm g} \Omega/ 2$), and simultaneously, 
scattering occurs three dimensionally.
Further, when the epicycle amplitude $e r_{\rm g} \Omega/\kappa$ is 
larger than the tidal radius $r_{\rm t}$, where $r_{\rm t}$ is defined by 
\eq
\begin{array}{lll}
\disp
r_{\rm t} & = & 
\disp
\left(\frac{G(m_1 + m_2)}{\alpha \Omega^2}\right)^{\frac{1}{3}} \\
\disp
    & = &
\disp
 r_{\rm g}\alpha^{-1/3},
\label{tidalradi}
\end{array}
\eqend
that is, when $e \ga (\kappa/\Omega) \alpha^{-1/3}$, orbits are not perturbed 
until the distance between bodies is much smaller than $r_{\rm t}$, 
since scattering cross section is small according to high relative velocity.
Fig.\ \ref{fig5}-c shows an example of orbital behaviour in this region.
The orbit is hardly perturbed except when the distance is well smaller than $r_{\rm t}$.
In this case, impulse approximation adopted by IKM93 is valid.
They adopted Rutherford scattering formula neglecting a disc potential, 
and assumed incident motion to the two-body Rutherford
scattering is described by 
the unperturbed motion given by Eqs.\ (\ref{xrelkai}) and (\ref{vrelkai}).
Actually, numerically obtained $P_{\rm heat}$, $Q_{\rm heat}$, 
and $e$ - $i$ ratio in this region are 
in good agreement with those given by IKM93:
in Fig.\ \ref{fig3}-c, 
dashed line denotes $p(e,i,b)$ and $q(e,i,b)$ calculated by IKM93, and 
the integrated quantities, i.e., $P_{\rm heat}$ and $Q_{\rm heat}$ are in agreement with those 
obtained by IKM93 within accuracy of 10 per cent.
In the region of small $b$, the analytical results
deviate from the numerical ones, however, this deviation disappears when
averaged over $b$.
This is because some cancelation with regard to $\tau$ or $\omega$
for fixed $b$ in the analytical calculation would be transferred to 
cancelation for slightly different $b$ by
weak distant perturbation in the numerical
calculation.

In Fig.\ \ref{fig4}, solid line denotes $ i/e = 0.68$ which is predicted by IKM93.
This is also consistent with numerical results for 
$e \ga (\kappa/\Omega) \alpha^{-1/3} \simeq 2.4$.

In the dispersion dominant region, $i/e$ in the equilibrium state is 
smaller than unity. 
The origin of this anisotropic velocity dispersion is explained as follows
(IKM93). 
In the dispersion dominant region, relative motion is approximated 
by unperturbed one as in Fig.\ \ref{fig5}-c, and close encounter 
always takes place when the
particle is moving leftward (rightward) if $b > 0$ ($b < 0$). 
Hence at the moment of the closest approach, the horizontal component of the particle's 
incident velocity is always decelerated by the shear motion.
On the other hand, the vertical motion is not affected by the shear 
and such a deceleration does not occur.
At the moment of the closest approach, two-body scattering neglecting 
the tidal force can be applied, so that scattering 
tends to make (local) velocity isotropic, that is, equal energy is 
partitioned to each direction locally.
Consequently, because of the deceleration of the horizontal motion 
at the closest approach, 
excessive energy is transferred to $x$- and $y$-directions compared to the 
$z$-direction.
Hence in the dispersion dominant region, 
the equilibrium value of $i/e$ is smaller than unity.

As shown in Fig.\ 3-c, heating is dominated almost equally by orbits 
with $ 0.6 \la b \la e \Omega/\kappa$. 
As described before, 
the upper limit $e \Omega/\kappa$ comes from the crossing condition of an 
unperturbed orbit. 
Hence, in large $e$ case, orbits with correspondingly  
large $b$ necessarily contribute to the 
heating. 
In other words, encounters with large shear necessarily contribute.
This is the case even in the limit of the solid-body rotation.
Thus IKM93 obtained $i/e \sim 0.8 (<1)$ even in the solid-body rotation case.
As shown below, however, this anisotropic dispersion region 
practically disappears. 

\vspace{0.5cm}
{\em horseshoe dominant region} 
\vspace{0.5cm}

In the region with $\alpha/\sqrt{2} \la e \la (\kappa/\Omega) \alpha^{-1/3}$, 
approach velocity is dominated by random velocity as in the dispersion dominant region.
However, orbital behaviour is quite different from that in the dispersion dominant 
region.
In this region, relative velocity is not as high as that in the dispersion dominant region,
so that relative motion is affected by distant perturbation
similar to the shear dominant region.
Figure 5-b shows an example of an orbit in this region. 
The orbit of guiding centre is gradually bent by Coriolis force until 
the orbit has $b$ of different sign to turn back.
Usually, such an orbit with $e = i = 0$ is 
called 'horseshoe orbits'.
Motion of guiding centre is similar to 'horseshoe orbits'
even in the case of $e, i \ne 0$, since
$e$ and $i$ are 
adiabatic invariants in distant region (H\'{e}non \& Petit 1986).
In this paper, we use the name 'horseshoe orbits' even 
if $e, i \neq 0$.

As shown in Fig.\ 3-b, heating is dominated by orbits with $0.6 \la b \la 2$ 
in this velocity region. 
When  $b \la 2$, the horseshoe orbits are common, because 
Coriolis force dominates shear motion. 
However, in contrast with shear-dominated case,
larger epicycle amplitude enables the bodies to closely approach.
In fact, close-encounting 'horseshoe orbits' as in Fig. 5-b
contribute a lot to $P_{\rm heat}$ and $Q_{\rm heat}$.

From the above orbital behaviour, the equilibrium ratio $i/e \sim 1$ is 
realized as follows.
On the contrary to the dispersion dominant case,
a close encounter always takes place when the particle is  
moving in the same direction as the guiding-centre motion 
(see Fig.\ \ref{fig5}-b).
Hence, the particle's horizontal motion is locally accelerated at the close encounter.
Consequently, the same argument as in the dispersion dominant region predicts that 
$i/e$ should became larger than unity.
However, because the motion of the guiding centre which is proportional to $b$ 
is relatively slow, significant anisotropy does not appear.
In the limit of the solid-body rotation, shear motion, and hence, 
the motion of 
the guiding centre slows down. Then  
$i/e \simeq 1$ would be realized.    

\vspace{0.5cm}

\begin{figure}
 \epsfbox{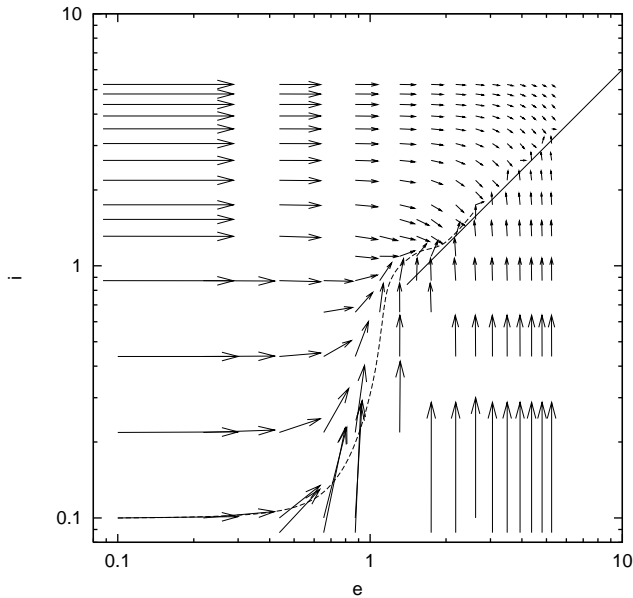}
 \caption{The same as Fig.\ 4 but $\kappa/\Omega = 1.58$.
 The solid line denotes $i = 0.60e$ predicted by IKM93}
 \label{fig6}
\end{figure}

\begin{figure}
 \epsfbox{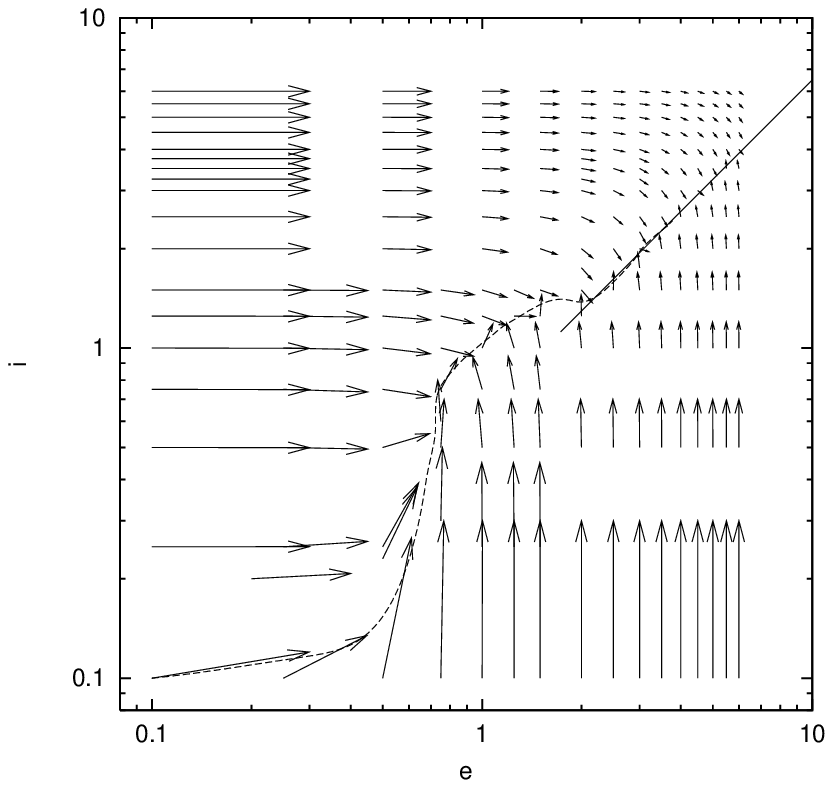}
 \caption{The same as Fig.\ 4 but $\kappa/\Omega = 1.73$.
 The solid line denotes $i = 0.65e$ predicted by IKM93}
 \label{fig7}
\end{figure}

So far we have considered the case where $\kappa/\Omega = 1.87$.
Our result in this case suggests that 
\eq
\left\{ \begin{array}{l}
\mbox{the shear dominant region} \mbox{:} \;\;\;\;\;\;\;\;\;\;\;\;\;\;\;\;
e \la \alpha/\sqrt{2}, \\
\mbox{the horseshoe dominant region} \mbox{:} \\
\;\;\;\;\;\;\;\;\;\;\;\;\;\;\;\;\;\;\;\;\;\;\;\;\;\;\;\;\;\;\;\;\;\;\;\;\;\;\; 
 \alpha/\sqrt{2} \la e \la (\kappa/\Omega)\alpha^{-1/3}, \\
\mbox{the dispersion dominant region}\mbox{:}\;\;\;
(\kappa/\Omega)\alpha^{-1/3} \la e,
\label{boundary}
\end{array}
\right.
\eqend
where $\alpha = 4 - \kappa^2/\Omega^2$.
To confirm the relation (\ref{boundary}), we also calculated the cases 
with $\kappa/\Omega =1.58$ and $1.73$ (Fig.\ \ref{fig6} and \ref{fig7}).
For $\kappa/\Omega = 1.58$, 
we calculated 208 sets of $(e,i)$, 
and for $\kappa/\Omega = 1.73$, 
218 sets of $(e,i)$ are calculated.
In each case, as in Fig.\ \ref{fig4}, we integrated  
typical trajectory of a particle on the $e$-$i$ plane (dashed lines).
The solid lines are also added as the result of IKM93.
Expected boundaries of the regions 
in the individual cases are shown in Table \ref{tab1}.
Figs. \ref{fig4}, \ref{fig6} and \ref{fig7} 
agree with Table \ref{tab1}.

\vspace{1em}
\begin{table}
\caption{Boundaries among three regions.}
\label{tab1}
\begin{tabular}{ccc} \hline
$\kappa/\Omega$ & $\alpha/\sqrt{2}$ & $(\kappa/\Omega)\alpha^{-1/3}$ \\ \hline\hline
1.58 & 1.0 & 1.4 \\
1.73 & 0.7 & 1.7 \\
1.87 & 0.35 & 2.4 \\ \hline
\end{tabular}
\end{table} 
\vspace{1em}

The horseshoe dominant region shrinks as $\kappa/\Omega$ decreases.
The horseshoe dominant region would vanish for 
$\alpha/\sqrt{2} \ga (\kappa/\Omega)\alpha^{-1/3}$ 
(equivalently, $\kappa/\Omega \la 1.5$). 
However, it should be noted that the 'horseshoe'-type close 
encounters still occur even if $\kappa/\Omega \la 1.5$.
We found when $\kappa/\Omega = 1.30$,  
this effect makes the equilibrium ratio slightly larger than that predicted by IKM93 for 
$e \sim 2$-$3$.

Since Ida (1990) and Kokubo \& Ida (1992) only studied 
the cases with $\kappa/\Omega = 1.0$ and $\kappa/\Omega = 1.39$, respectively, they 
did not find the horseshoe dominant region.
On the other hand, the horseshoe dominant region 
expands and dominates other two regions as $\kappa/\Omega$ approaches $2$ 
($\alpha \rightarrow 0$).
Therefore, in the limit $\kappa/\Omega \rightarrow 2$, the region
in which $i/e \sim 1$ is realized covers all over velocity space
except for $e \rightarrow \infty$.
In the limit with  $e \rightarrow \infty$, 
IKM93's analysis would still be correct.
The IKM93's analysis, which well accounts for $i/e$ in
low $\kappa/\Omega$ case, is also valid in 
high $\kappa/\Omega$ case if $e$ is sufficiently large, but the isotropic velocity region 
actually dominates in that case. 
Thus the contradiction stated in the introduction is solved. 

\subsection{The Effect of Averaging on the Rayleigh Distribution Function} 

Here we present the heating rates with the Rayleigh distribution. 
This is necessary not only because the realistic velocity distribution must be 
considered, but also because we compare the results to those of  
$N$-body simulations in the next section. 

\begin{figure}
 \epsfbox{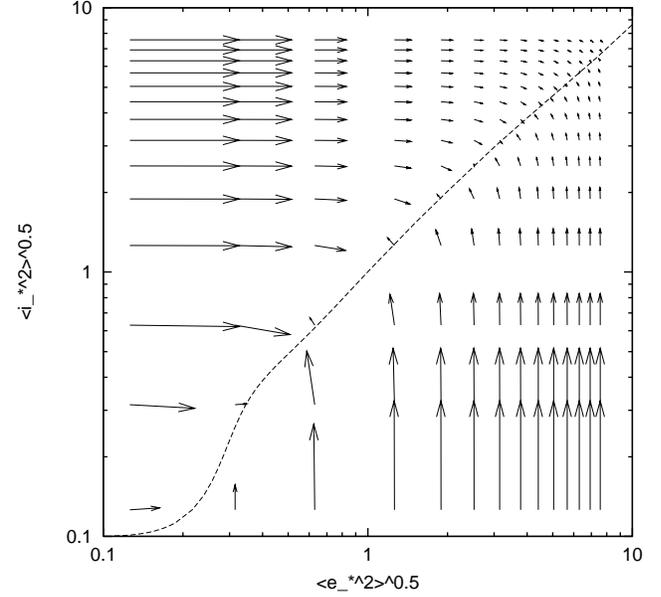}
 \caption{Direction of the evolution of $\ermsj$ and 
 $\irmsj$ plotted on the $\ermsj$ \mbox{-} $\irmsj$  
 plane in the case of $\kappa/\Omega = 1.87$.
 The angle and length of each vector is determined in the same way as 
 Fig.\ 4, but constants $a = 0.2$ and $b = 1000$.
 Similar to Figs.\ 4, 6, and 7, 
 we plot the typical trajectory as the dashed line.}
 \label{fig8}
\end{figure}

Our numerical calculation of the heating rates 
$P_{\rm heat}$ and $Q_{\rm heat}$ is restricted within $e$, $i \la 12$.
By comparing numerically obtained $P_{\rm heat}$ and $Q_{\rm heat}$ to 
those of IKM93, as stated in the last subsection, we find both are in agreement within 
accuracy of 10 per cent where $e, i \ga 7$ for $\kappa/\Omega = 1.87$.
Accordingly, in  
the region with $e \ga 12$ or $i \ga 12$, we use the analytical results of IKM93 in this case.  
In Fig.\ \ref{fig8}, we show the evolution diagram obtained from 
$\langle P_{\rm heat} \rangle$ and 
$\langle Q_{\rm heat} \rangle$ 
in the case of $\kappa/\Omega = 1.87$. 
The vectors drawn as a function of the root mean squares,
$\ermsj$ and $\irmsj$ in the same manner 
as in Fig.\ \ref{fig4} (here, $a = 0.2, b=1000$). 
Here, we use $\ermsj$ and $\irmsj$ as 
normalized
velocity dispersion of particles 
and distinguish them from those of relative motion, $\erms$ and $\irms$.
Note that $\ermsj$ ($\irmsj$) and $\erms$ ($\irms$) are related by Eqs.\ (\ref{erelation}).
When mass of test particles~1 is much smaller than that of field particles~2, 
$\ermsj = \erms$ and $\irmsj = \irms$ owing to energy equipartition. 
On the other hand, in the system of identical mass, $\ermsj = \erms/\sqrt{2}$ 
and $\irmsj = \irms/\sqrt{2}$.   
Hereafter we consider the system of identical mass in accordance with $N$-body simulations 
in the next section.
Hence it should be noted that $e_*$ is smaller than $e$ by a factor $\sqrt{2}$.
In this figure, we also plotted the typical trajectory in the same way as 
Fig.\ \ref{fig4}. 
Although averaging smoothes the boundaries of the three regions observed in Fig.\ \ref{fig4},  
the behaviour of the equilibrium ratio of the velocity dispersion 
is basically the same as Fig.\ \ref{fig4} :
In very small velocity case, $\ermsj \gg \irmsj$,
in intermediate velocity case ($\ermsj$ and $\irmsj$ are 
of order unity), the equilibrium value is almost unity, 
and in larger velocity case 
the equilibrium value is less than unity and seems to approach 
the value which predicted by IKM93
(for example, in the equilibrium state, $\irmsj / \ermsj \simeq 1.0$ for 
$\irmsj = 2.0$, $\irmsj / \ermsj \simeq 0.95$ for $\irmsj = 4.0$, 
and $\irmsj / \ermsj \simeq 0.86$ for $\irmsj = 9.0$).
It should be noted that the obtained equilibrium ratio seems to approach IKM93's 
much more moderately than the case of Fig.\ \ref{fig4}: 
In the case where the averaging is not done, our results (the equilibrium ratio) 
almost coincide with IKM93's for $e \ga 2$, while with the averaging, 
our results do not coincide with IKM93's even when $\ermsj$, $\irmsj \sim 10$. 
In other words, the horseshoe dominant region influences even if $\ermsj$ and $\irmsj$ are 
much larger than unity as a result of the averaging on the velocity distribution. 
Actually, we found the influence of the horseshoe dominant region remains as long as 
$\irmsj \la 30$ 
in the case of $\kappa/\Omega = 1.87$ and $\irmsj \la 20$ for $\kappa/\Omega = 1.58$. 
On the other hand, in the case of Kepler rotation, where there is no horseshoe dominant
region, we found that our numerical results coincide with IKM93's for $\irmsj \ga 2$.  

\section{$N$-BODY SIMULATIONS OF THE DISC HEATING}

\subsection{Basic equations}

In section~2, we studied disc heating  
through statistical compilation of independent two-body scatterings.
In this section, to check
the results in section~2, 
we perform direct $N$-body simulations of particles in
axisymmetric disc potentials.
We consider $N$ self-gravitating particles revolving  in a disc potential.
Then the motion of each particle is described as
\begin{equation}
\frac{{\rm d}\bmath{v_j}}{{\rm d}t} = 
\sum_{i \neq j}^{N} G m_j 
\frac{\bmath{x_i} - \bmath{x_j}}
{| \bmath{x_i} - \bmath{x_j} |^3}
+ \bmath{F_j}  \mbox{ ~ }(j = 1,\ldots ,N),
\label{eq311}
\end{equation}
where the subscript $j$ indicates particle's index,
$\bmath{v_j}$, $\bmath{x_j}$, and $m_j$ are the velocity vector, the 
position vector, and the mass of the particle $j$.
The last term on the right hand side, $\bmath{F_j}$, is 
external force resulted from the disc potential. 
We consider the radial and vertical excursion of each particle are 
sufficiently small compared to its orbital radius
in accordance with the study in the last section. 
In this case, the external force field $\bmath{F_j}$ is expressed 
by two parameters of the disc potential, $\alpha$ and $\nu/\Omega$, as
\eq
\left\{
\begin{array}{l}
\disp
F_{x j} = -a \Omega^2 \left(\frac{x_j}{a}\right)
\disp
\left[ \frac{R_j^2}{a^2} + \left(\frac{\nu}{\Omega}\right)^2 \frac{z_j^2}{a^2} 
\disp
\right]^{-\frac{\alpha}{2}}, \\
\disp
F_{y j} = -a \Omega^2 \left(\frac{y_j}{a}\right)
\disp
\left[ \frac{R_j^2}{a^2} + \left(\frac{\nu}{\Omega}\right)^2 \frac{z_j^2}{a^2} 
\disp
\right]^{-\frac{\alpha}{2}}, \\
\label{eq312}
\disp
F_{z j} = -a \Omega^2 \left[\left(\frac{\nu}{\Omega}\right)^2 \frac{z_j}{a}\right]
\disp
\left[ \frac{R_j^2}{a^2} + \left(\frac{\nu}{\Omega}\right)^2 \frac{z_j^2}{a^2} 
\disp
\right]^{-\frac{\alpha}{2}}, 
\end{array}
\right.
\eqend
where Cartesian coordinates $(x, y, z)$ is centred at the bottom of the 
disc potential, 
$R$ is the radial component of cylindrical coordinates given by
$ R = (x^2 + y^2)^{1/2} $.

We integrate Eqs.\ (\ref{eq311}) by using the fourth-order Hermite scheme
with the individual and hierarchical timestep (Makino 1991). 
The most expensive part of the Hermite scheme is 
the calculation of the force and its time-derivative 
whose cost increases in proportion to $N^2$ because we calculate the direct sum of all pairs.
To reduce the computational time of this part, we employed
a special purpose hardware, HARP-2 (Makino, Kokubo \& Taiji 1993)
and GRAPE-4 (Makino, Taiji, Ebisuzaki \& Sugimoto 1997).

\subsection{Initial conditions of the swarm of particles}

We mainly investigated the $\kappa/\Omega$ 
dependence of the equilibrium ratio of velocity dispersion.
The parameter $\nu/\Omega$ is fixed to be unity in most cases, since 
IKM93 suggested that it does not affect the equilibrium ratio
(we also did some $N$-body simulations with $\nu/\Omega \neq 1$ and found 
that the ratio is the same as in the case with $\nu/\Omega = 1$).
We did 27 $N$-body simulations with nine different values of 
$\kappa/\Omega$.
For each value of $\kappa/\Omega$, we did several runs starting with 
different initial conditions.

In Table \ref{tab2}, we summarize the initial conditions.
We distribute 1000 particles with identical mass
randomly in the region 
$a - \Delta a/2 < a < a + \Delta a /2$.
In most cases, the particle masses are 
$m = 3 \times 10^{-9} M_{\rm g}$ where $M_{\rm g}$
is effective mass of the center defined by $GM_{\rm g} / a^2 = \Omega^2 a$.
As suggested by the argument in section~2, 
the results would be independent of particle masses, if the results
are scaled by $r_{\rm g}$.
We took $\Delta a \ll a$ to make simulation radially local, 
however, we took $\Delta a$ sufficiently larger than $r_{\rm t}$ 
(characteristic size of a particle's potential well) and 
epicycle amplitude for the edge effects to be negligible. 
We used enough number ($N=1000$) of bodies that 
the bodies can closely approach each other.
Initially, we set the same $e_*$ and $i_*$ $(e_{*0},i_{*0})$  for all particles,
however, the velocity distribution
converges to the Rayleigh distribution given by Eq.\ (\ref{Rayleigh}) 
in shorter time interval compared to the characteristic time-scale of 
the evolution of the velocity dispersion (i.e., two-body relaxation time $T_{\rm 2B}$). 

\begin{table}
\caption{Initial parameters in $N$-body simulations. }
\label{tab2}
\begin{tabular}{cccccc}\hline
run & $\kappa/\Omega$ & $\nu/\Omega$ & $e_{*0}$ & $i_{*0}$ & $\Delta a /a$ 
  \\ \hline \hline 
1 & 1.00 & 1.00 & 0.6 & 0.6 & 0.29\\
2 & 1.00 & 1.00 & 0.6 & 1.2 & 0.27\\ 
3 & 1.00 & 1.00 & 1.1 & 0.3 & 0.27\\ \hline

4 & 1.10 & 1.00 & 0.2 & 0.2 & 0.13\\
5 & 1.10 & 1.00 & 0.2 & 0.4 & 0.13\\ \hline

6 & 1.20 & 1.00 & 0.5 & 0.5 & 0.27\\
7 & 1.20 & 1.00 & 0.5 & 0.8 & 0.27\\ \hline
 
8 & 1.30 & 1.00 & 0.6 & 0.6 & 0.33\\
9 & 1.30 & 1.00 & 0.6 & 1.2 & 0.27\\
10 & 1.30 & 1.00 & 1.2 & 0.3 & 0.27\\ 
11 & 1.30 & 2.00 & 0.6 & 0.6 & 0.29\\
12 & 1.30 & 4.00 & 0.6 & 0.6 & 0.29\\ \hline

13 & 1.58 & 1.00 & 2.0 & 1.0 & 0.27\\ 
14 & 1.58 & 1.00 & 1.0 & 2.0 & 0.27\\ \hline 

15& 1.73 & 1.00 & 0.1 & 0.1 & 0.27\\
16& 1.73 & 1.00 & 0.4 & 0.2 & 0.27\\ \hline

17& 1.80 & 1.00 & 1.0 & 0.5 & 0.27\\
18& 1.80 & 1.00 & 2.0 & 1.0 & 0.27\\
19& 1.80 & 1.00 & 1.0 & 2.0 & 0.27\\ \hline

20& 1.87 & 1.00 & 1.1 & 0.6  & 0.27\\
21& 1.87 & 1.00 & 2.0 & 1.0  & 0.27\\
22& 1.87 & 1.00 & 1.0 & 2.0  & 0.27\\ 
23& 1.87 & 2.00 & 1.0 & 2.0 & 0.27\\ 
24& 1.87 & 4.00 & 1.0 & 2.0 & 0.27\\ \hline

25& 1.95 & 1.00 & 1.0 & 0.5  & 0.27\\ 
26& 1.95 & 1.00 & 2.0 & 1.0  & 0.27\\ 
27& 1.95 & 1.00 & 1.0 & 2.0  & 0.27\\ \hline
\end{tabular}
\end{table}

\subsection{Results of the $N$-body simulations}

First we show the detailed results in two characteristic cases 
with $\kappa/\Omega = 1.30$ and $\kappa/\Omega = 1.87$.
As stated before, orbital properties change near 
$\kappa/\Omega \sim 1.5$.

\begin{figure}
 \epsfbox{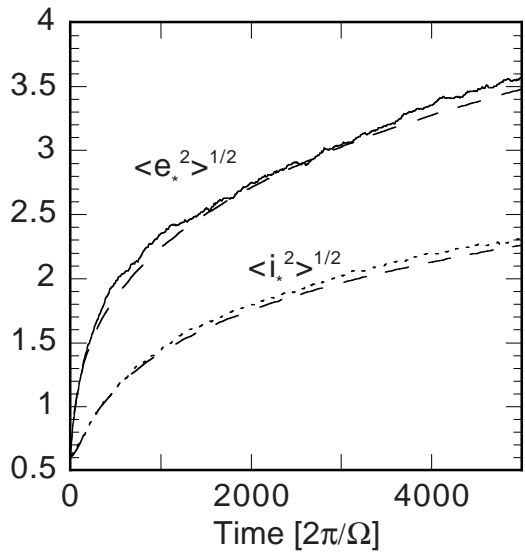}
 \caption{Time evolutions of $\ermsj$ (solid line) and 
$\irmsj$ (dotted line) in the case of run 8.
Time is scaled by the rotation period at $R = a$ ($2\pi/\Omega(a)$).	
The dashed lines denote time evolutions of the 
normalized
velocity dispersion predicted by 
${\rm d}\langle e_*^2 \rangle/{\rm d}t$ and ${\rm d} \langle i_*^2 \rangle/{\rm d}t$ 
given by Eqs.\ (13) and the two-body results in section~2.} 
\label{fig9}
\end{figure}

\begin{figure}
 \epsfbox{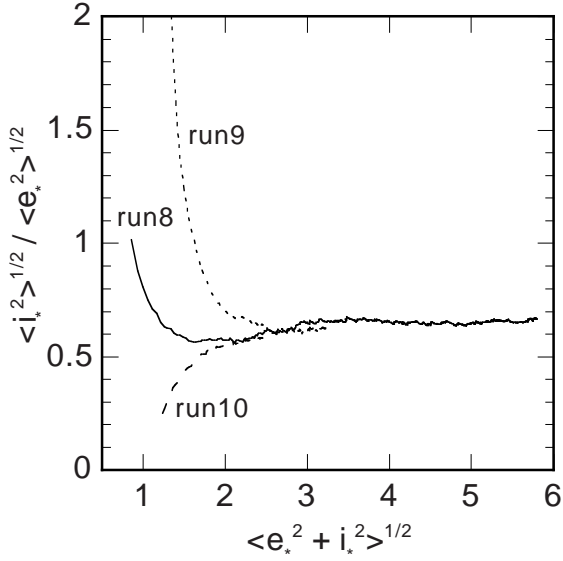}
 \caption{The evolution of ratio $\irmsj / \ermsj$ against 
$\langle e_*^2 + i_*^2 \rangle^{1/2}$ in the cases of run 8, 9, and 10
(the same initial parameters but different sets of initial $e_{*0}$ and $i_{*0}$).
This is equivalent to the time evolution, since 
$\langle e_*^2 + i_*^2 \rangle^{1/2}$
monotonically increases as time evolves.}
 \label{fig10}
 \end{figure}

In Fig.\ \ref{fig9}, we show the time evolutions of $\ermsj$ (solid line) 
and $\irmsj$ (dotted line) of run 8 ($\kappa/\Omega = 1.30$).
The parameter $\kappa/\Omega = 1.30$ corresponds to
the galactic potential in the solar neighbourhood.
In the figure, time is scaled by the rotation period at $a$, 
namely, $2\pi/\Omega(a)$.
Although both $\ermsj$ and $\irmsj$ keep growing,
the ratio of $\irmsj$ to $\ermsj$ seems to be constant
in the later stage.
The ratio $\irmsj / \ermsj$ is plotted as a function of
$\langle e_*^2 + i_*^2 \rangle^{1/2}$ in Fig.\ \ref{fig10}, 
which is equivalently, time evolution of $\irmsj / \ermsj$.
We also plot the results with different sets of $(e_{*0}, i_{*0})$
for the same $\kappa/\Omega$ (run 9 and 10).
The ratio of the velocity dispersions settles to a certain constant 
($\sim 0.6$) value independent of initial values of $e_*$ and $i_*$ after 
$\langle e_*^2 + i_*^2 \rangle^{1/2}$ exceeds about 3.
The equilibrium ratio gradually increases as $\langle e_*^2 + i_*^2 \rangle^{1/2}$,
which is consistent with the analytical argument in IKM93 that
the equilibrium ratio has a weak
dependence on $\langle e_*^2 + i_*^2 \rangle^{1/2}$.

The equilibrium ratio $\irmsj / \ermsj \sim 0.6$ is consistent with the 
observational value given by Wielen (1977) and Chen et al.\ (1996), 
statistical compilation of two-body encounters by Kokubo \& Ida (1992),
and $N$-body simulation by Villumsen (1985).

\begin{figure}
 \epsfbox{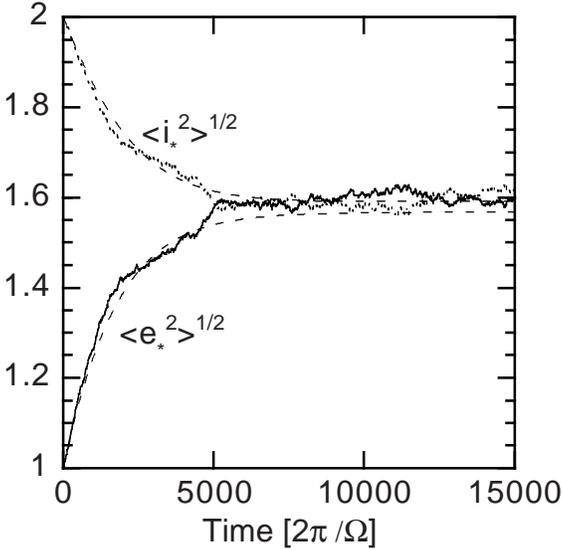}
 \caption{Time evolutions of $\ermsj$ (solid line) and $\irmsj$ (dotted line) 
in the case of run 22.
The dashed lines denote time evolutions of the 
 normalized
velocity dispersion predicted by 
the two-body results in section~2.}
 \label{fig11}
\end{figure}

The time evolution of $\ermsj$ and $\irmsj$ 
in the case of $\kappa/\Omega = 1.87$ 
(run 22) is shown in Fig.\ \ref{fig11}.
Solid and dotted lines express $\ermsj$ and $\irmsj$.
The increase in $\ermsj$ is almost compensated by decrease in $\irmsj$.
The system evolve to the state of the equilibrium ratio, 
without increase in $\langle e_*^2 + i_*^2 \rangle^{1/2}$, which is in contrast to
the case of $\kappa/\Omega = 1.30$.
Actually, the normalized random velocity $\langle e_*^2 + i_*^2 \rangle^{1/2}$
only increases about 1 per cent throughout the simulation.
This is consistent with the argument given in section~2 and appendix~A
($T_{\rm ratio} \ll T_{\rm random}$).
In the potential with $\kappa/\Omega = 1.87$, orbital motion is
close to solid-body rotation.
Since disc heating is caused by transformation from shear motion to
random motion, the disc heating is weak in the potential 
with $\kappa/\Omega = 1.87$.
Gravitational interactions mostly result in redistribution of random energy.

\begin{figure}
 \epsfbox{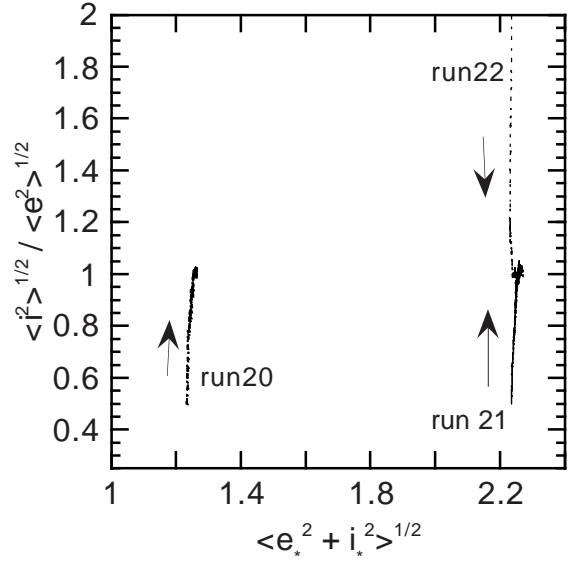}
 \caption{The evolution of ratio $\irmsj / \ermsj$ against 
$\langle e_*^2 + i_*^2 \rangle^{1/2}$ in the cases of run 20, 21, and 22.}
 \label{fig12}
\end{figure}

\begin{figure}
 \epsfbox{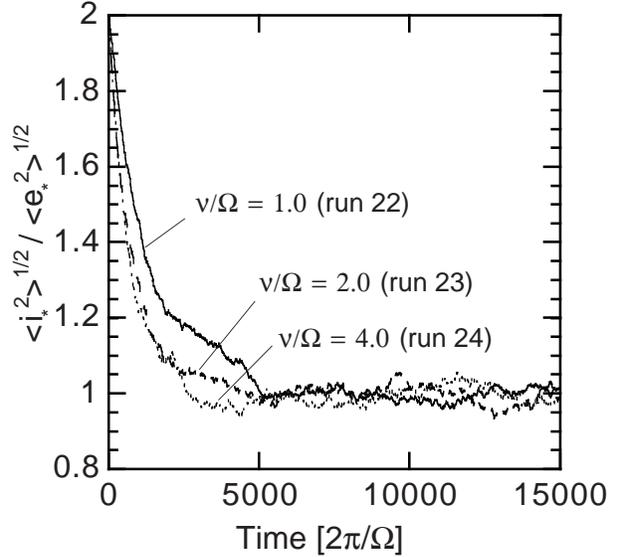}
 \caption{The time evolution of ratio $\irmsj/\ermsj$ with various $\nu/\Omega$.
Solid line corresponds to the results of run 22 ($\nu/\Omega = 1.0$).
Dashed and dotted lines for run 23 ($\nu/\Omega = 2.0$) and 
run 24 ($\nu/\Omega = 4.0$).}
 \label{fig13}
\end{figure}

In Fig.\ \ref{fig12}, we show the evolution of 
$\irmsj/\ermsj$ with different values of $\langle e_*^2 + i_*^2 \rangle^{1/2}$
(run 20, 21 and 22).
This figure shows the equilibrium ratio settles 
to nearly unity independent of 
the initial conditions, as long as $\ermsj$ and $\irmsj$ 
are of order unity.

We found that the distributions of the $e_*$ and $i_*$ 
evolve from the initial $\delta$-function type function to 
the Rayleigh distribution in a time interval about  
$< 0.1 T_{\rm 2B}$ in both cases.
Hence the assumption in Eq.(\ref{Rayleigh}) is valid.

In Figs.\ \ref{fig9} and \ref{fig11}, we also plotted the evolution of the 
normalized
velocity dispersion
calculated from ${\rm d}\langle e_*^2 \rangle /{\rm d}t$ and 
${\rm d}\langle i_*^2\rangle/{\rm d}t$ obtained in section~2 (Eqs.\ (\ref{evolution6})) as 
dashed lines.
In both cases, the predicted results are in good agreement with 
the results of $N$-body simulations. 

\begin{figure*}
 \epsfbox{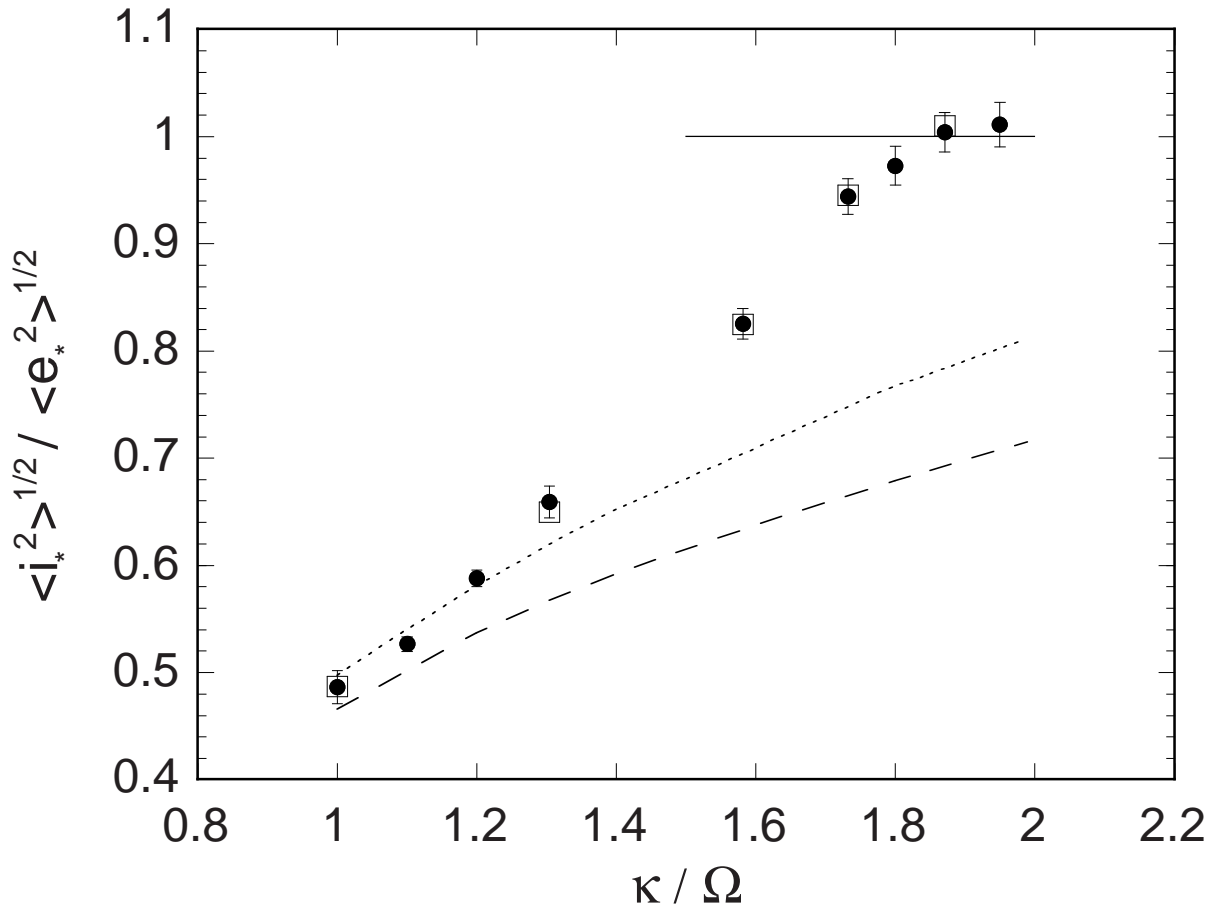}
 \caption{Equilibrium value of $\irmsj / \ermsj$ as a function of 
$\kappa/\Omega$. 
Points with error bars are the results of the 27 $N$-body simulations.
Initial conditions of these simulations are presented in Table 2.
Solid line corresponds to the isotropic state, i.e., $\irmsj / \ermsj = 1.0$,
which is realized in the horseshoe dominant region.
Dashed and dotted curves denote the results of IKM93 in the cases where 
 $\irmsj = 1.5$ and $\irmsj = 5.0$, respectively. 
Open squares are the equilibrium ratio obtained by 
the two-body results in section~2
at the mean velocity corresponding to the $N$-body results.}
 \label{fig14}
\end{figure*}

In addition to the runs with $\nu/\Omega = 1.0$, we also did 4 simulations 
with $\nu/\Omega = 2.0$ and $4.0$ (run 11, 12, 23, and 24).  
In the solar neighbourhood, 
$\nu/\Omega \simeq 4$ (Binney \& Tremaine 1987).
In these simulations, 
the other initial conditions are the same as run 8 for run 11 and 12, 
and the same as run 22 for run 23 and 24.  
The time evolution of $\irmsj/\ermsj$ in the case of $\kappa/\Omega 
= 1.87$ is shown in Fig.\ \ref{fig13}.
The solid line denotes the result of run 22 ($\nu/\Omega = 1.0$), 
the dashed line is that for run 23 ($\nu/\Omega = 2.0$), 
and the dotted line for run 24 ($\nu/\Omega = 4.0$).
The variation of $\nu/\Omega$ does not 
affect the equilibrium ratio as expected, though the larger $\nu/\Omega$ 
results in the faster relaxation to the equilibrium state 
(This is because the larger $\nu/\Omega$ leads to smaller disc scale height 
and therefore leads to more frequent scatterings among particles: 
see Eqs.\ (\ref{eq2120})).
These results are the cases also for $\kappa/\Omega = 1.30$.

We also carried out $N$-body simulations in the 
cases of $\kappa/\Omega = 1.0$, $1.1$, $1.2$, $1.73$, and $1.95$, 
as summarized in Table \ref{tab2}.
In Fig.\ \ref{fig14}, we plot the equilibrium value of $\irmsj/\ermsj$ 
obtained by the $N$-body simulations with filled circles where
error bars indicate 
standard deviation
in time evolution.
In the shear dominant region, we found timescale for heating 
(i.e. $T_{\rm 2B}$) is comparable to or shorter than that for change of  
the ratio. 
Thus in the shear dominant region, there is no 'equilibrium' ratio.
Accordingly,
we are interested in the ratio in 
the dispersion dominant and the horseshoe dominant regions not in 
the shear dominant region.
Since we cannot calculate the region with very large $\ermsj$ and $\irmsj$
for cpu limit,
we only plot the results in the horseshoe dominant region in 
the cases of $\kappa/\Omega \ga 1.60$.
In the cases of $\kappa/\Omega \la 1.30$, we plot the ratio
in the dispersion dominant region, since there is no horseshoe dominant region.

Together with the $N$-body results, 
the line $\irmsj/\ermsj = 1$ (the solid line) and
the analytic lines in the dispersion dominant region by IKM93 
are plotted. The dashed and dotted lines are IKM93's results with $\irmsj=1.5$ and $5.0$.
In the plotted $N$-body results, $\irmsj \simeq 1.5$-$5$ except
run 15, 18, and 21.  

\begin{figure*}
 \epsfbox{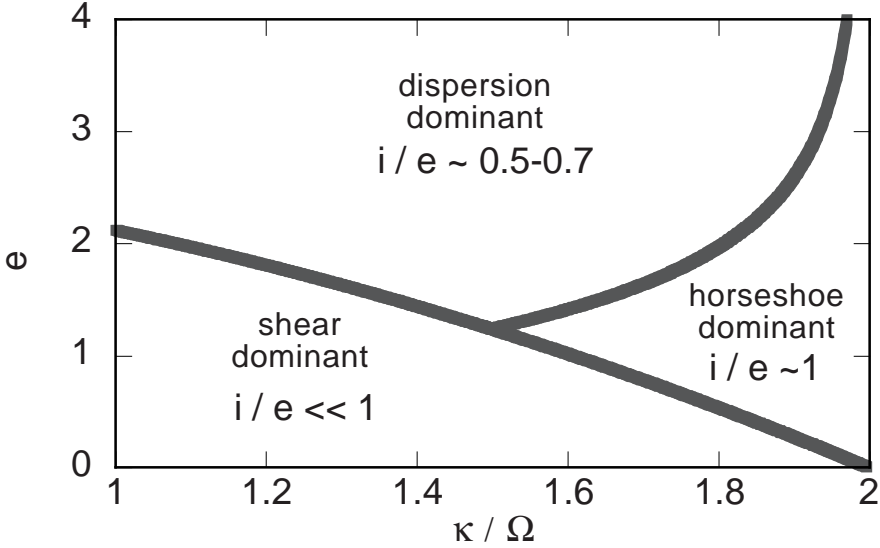}
 \caption{Boundaries among three regions obtained from the relation 
(25) and corresponding equilibrium ratios.}
 \label{fig15}
 \end{figure*}

For $\kappa/\Omega \sim 1.0$-$1.2$, the $N$-body results almost agree with
the analytical line. 
For $\kappa/\Omega \ga 1.87$, the $N$-body results show
isotropy ($\irmsj/\ermsj \simeq 1$), which is expected in
the horseshoe dominant region.  
However, for $1.6 \la \kappa/\Omega \la 1.8$, 
the $N$-body results are between
the lines of IKM93 and the isotropy.
As shown in section~2, the horseshoe dominant region is 
in limited velocity space in the case of relatively small $\kappa/\Omega$.
Since the velocity distribution (Rayleigh distribution)
includes ``dispersion-dominant'' encounters more or less,
its effect decreases the $\irmsj/\ermsj$ from 1.
On the other hand, the contamination of 'horseshoe'-type encounter
would make $\irmsj/\ermsj$ slightly larger than IKM93's result
as suggested in section~2. 
To confirm this interpretation, in Fig.\ \ref{fig14} we also plot
the results (squares) of section~2 with Rayleigh distribution
at the mean velocity corresponding to the $N$-body results.
They are consistent with the $N$-body results.
It is expected that numerical results would be consistent with IKM93's 
in sufficiently high mean velocity cases, because the effect of the 
horseshoe dominant region diminishes then.
In section~2, however, we suggested influence of the horseshoe dominant region remains
when $\irmsj <  20$-$30$ in the case where $\kappa/\Omega = 1.58$-$1.87$,
as a result of the averaging effect.
Unfortunately, because of numerical difficulty we cannot directly examine such a high velocity
region with neither $N$-body simulation nor the method in section~2.
 $N$-body simulation with very high random velocity needs 
considerable cpu time since two-body relaxation time is proportional 
to the forth power of the random velocity in a disc system (Appendix A).
The method in section~2 is difficult to apply since 
scattering cross-section becomes very small so that 
we cannot obtain reliable results.
In the limit of $\kappa/\Omega \rightarrow 2.0$,
the effect of horseshoe dominant encounters would always dominate
and $N$-body simulations always show $\irmsj/\ermsj \simeq 1$.

In summary, the results of $N$-body simulation are consistent with
the results in section~2.
Therefore, the physical argument in section~2 (mainly with
quantities before velocity averaging, $P_{\rm heat}$ and 
$Q_{\rm heat}$) would be valid. 

\section{CONCLUSIONS}

We have studied the ratio of the velocity dispersion 
($\sigma_z / \sigma_R$) which is attained 
through
mutual gravitational interaction among bodies in disc potentials.
We examined the cases with Kepler rotation $\kappa/\Omega=1$
to solid-body rotation $\kappa/\Omega=2$, 
where $\kappa$ and $\Omega$ are the epicycle and circular frequencies.
Another parameter $\nu/\Omega$ does not affects the result, 
because shear motion of particles, which is the origin of the anisotropic 
velocity dispersion, is independent of $\nu/\Omega$.  
We employed two different numerical methods, 
statistical compilation of two-body
encounters (section~2) and $N$-body simulations (section~3).
With the former method the physical properties are clearer and
wider parameter range can be examined, while the results are not
direct and some assumptions are introduced in the statistical
compilation.
On the other hand, the latter method is direct, although parameter range
we can simulate is restricted by cpu time.
The combination of the complementary methods would enable us to derive conclusive
results.

We found that the ratio becomes the equilibrium state
much more quickly than the amplitude of the velocity
dispersion changes except when $\kappa/\Omega$ is near 1.
The equilibrium ratio depends on amplitude of velocity
dispersion and disc potential parameter, $\kappa/\Omega$.
We found three characteristic velocity regimes:
\eq
\left\{ \begin{array}{l}
\mbox{the shear dominant region} \mbox{:} \;\;\;\;\;\;\;\;\;\;\;\;\;\;\;\;
e \la \alpha/\sqrt{2}, \\
\mbox{the horseshoe dominant region} \mbox{:} \\
\;\;\;\;\;\;\;\;\;\;\;\;\;\;\;\;\;\;\;\;\;\;\;\;\;\;\;\;\;\;\;\;\;\;\;\;\;\;\; 
 \alpha/\sqrt{2} \la e \la (\kappa/\Omega)\alpha^{-1/3}, \\
\mbox{the dispersion dominant region}\mbox{:}\;\;\;
(\kappa/\Omega)\alpha^{-1/3} \la e,
\label{boundary2}
\end{array}
\right.
\eqend
where $\alpha = 4 - \kappa^2/\Omega^2$ and $e$ corresponds to the amplitude of the random velocity of the relative motion of two particles
normalized by $r_{\rm g} \Omega$.
The velocity dispersion $\sigma_R$ and $e$ are related as 
$\erms = 2\sigma_R/(r_{\rm g} \Omega)$ for a system of identical particle 
or $\erms = \sqrt{2}\,\sigma_R/(r_{\rm g} \Omega)$ for a system with 
large mass ratio such as stars and 
giant molecular clouds (see Eqs.\ (\ref{defsigma}) and (\ref{erelation})).
The characteristic radius $r_{\rm g}$ is defined by Eq.(\ref{hillradius})
and it is related to tidal radius $r_{\rm t}$ as $r_{\rm g}=\alpha^{1/3}r_{\rm t}$.  

In the shear dominant region, shear motion dominates 
epicycle motion. 
Since shear motion is horizontal and orbits are bent before
the bodies come close to each other,
gravitational scattering takes place two-dimensionally. 
As a result, only $\sigma_R$ is raised so that $\sigma_z / \sigma_R \ll 1$.
In the dispersion dominant region,
epicycle velocity is so large that 
orbits are not perturbed until they closely approach each other.
The close encounters are well approximated by Rutherford formula
neglecting tidal force, as Lacey(1984) and IKM93 did.
As explained in section~2,
energy equipartition in horizontal and vertical motion at the
close encounters and the deceleration in horizontal velocity
by shear motion lead to excessive $\sigma_R$.
IKM93 predicted $\sigma_z/\sigma_R = 0.5$-$0.8$ ($\sigma_z/\sigma_R$ is
smaller for
smaller $\kappa/\Omega$).
Our numerical simulations agree with IKM93's prediction
for relatively small $\kappa/\Omega$ and suggest
agreement even for larger $\kappa/\Omega$. 
However, dispersion dominant region,
where IKM93's prediction is valid, is overwhelmed by
the newly found horseshoe dominant region in velocity space
in the case of $\kappa/\Omega \sim 2$ (See (\ref{boundary2})).
In the horseshoe dominant region, 'horseshoe'-type close encounters
dominates gravitational relaxation.
In this case, $\sigma_z/\sigma_R \sim 1$ is predicted.
The physical reason for $\sigma_z/\sigma_R \sim 1$ is given in
section~2.
Due to the contamination from the encounters in other velocity
regimes (the particles velocities have Rayleigh distribution),
$N$-body simulations usually show $\sigma_z/\sigma_R$ smaller than 1.
However, $\sigma_z/\sigma_R \sim 1$ is actually shown in
the case of $\kappa/\Omega \sim 2$ as in Fig.\ \ref{fig14} 
because the contamination diminishes as $\kappa/\Omega \rightarrow 2$.


In Fig.\ \ref{fig15}, we draw a schematic figure on
the ratio of velocity dispersion of self-gravitating
particles in disc potentials from Kepler rotation to solid-body rotation.
The boundaries of the three regions are 
given by the relation (\ref{boundary2}). 
In reality, the velocity distribution of particles makes the boundaries obscured
through the averaging on the Rayleigh distribution.

\section*{ACKNOWLEDGMENTS}

We thank Hiroyuki Emori and Hidekazu Tanaka for fruitful discussion and
for useful comments on the numerical codes.
We also thank Kiyoshi Nakazawa for continuous encouragement.

\appendix
\section{}

\begin{figure}
 \epsfbox{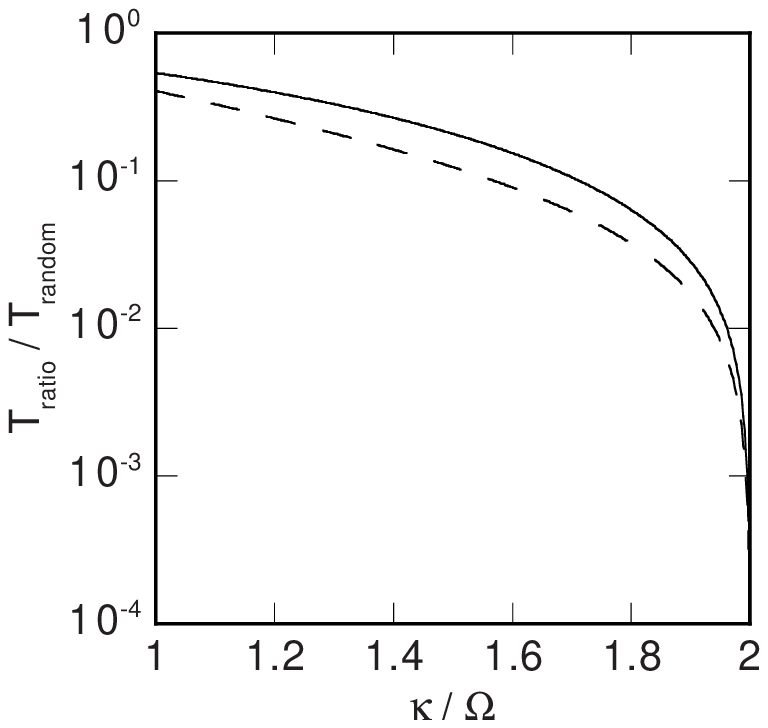}
 \caption{The ratio between $T_{\rm ratio}$ and $T_{\rm random}$ 
against $\kappa/\Omega$.
The solid line and the dashed line correspond to $i/e = 1.0$ and $2.0$.}
 \label{figA1}
\end{figure}

We evaluate the time-scale for the equilibrium state 
of $i/e$ to be realized and that for the random energy ($e^2 + i^2$)  
to be increased in the dispersion dominant region.
The former is defined by 
\eq
T_{\rm ratio} = \left[\frac{e}{i}\frac{{\rm d}(i/e)}{{\rm d}t}\right]^{-1},
\label{tratio}
\eqend
and the latter is 
\eq
T_{\rm random} = \left[\frac{1}{e^2+i^2}\frac{{\rm d}(e^2+i^2)}{{\rm d}t}\right]^{-1}.
\label{trandom}
\eqend
Accordingly, $T_{\rm ratio}$ and $T_{\rm random}$ are 
obtained from change rates of $e^2$ and $i^2$.
When $e$ and $i$ are large enough that the impulse approximation  
is valid, they are written as (Ida et al. 1993, Tanaka \& Ida 1996):
\eq
\left\{
\begin{array}{lc}
\disp 
\frac{{\rm d}e^2}{{\rm d}t} = C 
\disp
\left[ (1+\xi^2)K(\lambda) - \frac{3e^2}{i^2+\xi^{-2}e^2}E(\lambda)\right],\\
\label{heatappend}
\disp
\frac{{\rm d}i^2}{{\rm d}t} = C 
\disp
\left[ K(\lambda) - \frac{3i^2}{i^2+\xi^{-2}e^2}E(\lambda)\right],
\end{array}
\right.
\eqend
where $K$ and $E$ are the complete elliptic integral of the first and 
second kind, $\xi = 2\Omega/\kappa$, and 
$\lambda^2 = (1-\xi^{-2})e^2 /(e^2 + i^2)$.
The factor $C$ is defined by 
\eq
C = \frac{4}{\pi}\frac{\nu}{\Omega}n_s 
\disp
\left[\ln(1+\Lambda^2) - \frac{\Lambda^2}{1+\Lambda^2}\right]
\disp
\frac{1}{i \sqrt{e^2+i^2}}
\eqend
From Eqs.\ (\ref{tratio}), (\ref{trandom}), and (\ref{heatappend}), 
We can rewrite $T_{\rm ratio}$ and $T_{\rm random}$ as
\eq
\left\{
\begin{array}{ll}
\disp
T_{\rm ratio}  =  \disp \frac{e^2+i^2}{C}
\disp
\left\{\left[\frac{e}{i}-\frac{i}{e}(1+\xi^2)\right] K(\lambda) \right\}^{-1},\\
\disp
T_{\rm random}  =  \disp \frac{e^2+i^2}{C}  \\
\disp
 \times \left[(2+\xi^2)K(\lambda) 
\disp
- 3\left(\frac{e^2+i^2}{i^2+\xi^{-2}e^2}\right)E(\lambda)\right]^{-1}.
\end{array}
\right.
\eqend
In above equations, when the non-dimensional factors multiplied to 
$(e^2+i^2)/C$ 
are order unity, it is easy to see both $T_{\rm ratio}$ and $T_{\rm random}$
are the same order as the Chandrasekhar's two body relaxation time.
In the case where the disc potential is close to that of solid-body 
rotation ($\xi \rightarrow 1$ or $\kappa/\Omega \rightarrow 2.0$), 
however, $T_{\rm random}$ tends toward infinity,
while $T_{\rm ratio}$ does not (we consider the case where particles have not 
reach the state of the equilibrium ratio yet).
In Fig.\ \ref{figA1}, 
$T_{\rm ratio}/T_{\rm random}$ as a function of $\kappa/\Omega$ 
is plotted 
in the cases where $i/e = 1.0$ and $2.0$ (note that $T_{\rm ratio}/T_{\rm random}$
is a function of $i/e$, but not $e$ nor $i$).
When $\kappa/\Omega$ is nearly 1 ($\xi \sim 2$), 
$T_{\rm ratio}$ and $T_{\rm random}$ are the same order.
On the other hand, when $\kappa/\Omega$ is nearly $2$ ($\xi \sim 1$), 
$T_{\rm random}$ is much larger than $T_{\rm ratio}$.
In the latter case, 
disc heating proceeds quasi-stationarily compared to the process to reach 
the state of the equilibrium ratio.   

\begin{thebibliography}{}
\bibitem[\protect\citename{Barbanis \& Woltjer}1967]{BW}
Barbanis B., Woltjer L., 1967, ApJ, 150, 461
%
\bibitem[\protect\citename{Binney \& Lacey}1988]{BL88}
Binney J., Lacey C., 1988, MNRAS, 230, 597
%
\bibitem[\protect\citename{Binney \& Tremaine}1987]{gd}
 Binney J., Tremaine S., 1987, Galactic Dynamics.
Prinston Univ. Press, Prinston, New Jersey.
%
\bibitem[\protect\citename{Brown}1911]{B11}
Brown E. W., 1911, MNRAS, 71, 438
%
\bibitem[\protect\citename{Chandrasekhar}1949]{sd}
Chandrasekhar S., 1949, Principles of Stellar Dynamics.
Yale Univ. Press, New Haven, CN.
%
\bibitem[\protect\citename{Chen et al.\ }1997]{C97}
Chen B., Asiain R., Figueras F., Torra J., 1997, A\&A, 318, 29
%
\bibitem[\protect\citename{Emori et al.\ }1993]{E93}
Emori H., Ida S., Nakazawa K., 1993, PASJ, 45, 321
%
\bibitem[\protect\citename{Hesegawa \& Nakazawa}1990]{HN90}
Hasegawa M., Nakazawa K., 1990, A\&A, 227, 619
%
\bibitem[\protect\citename{H\'{e}non \& Petit}1986]{HP86}
H\'{e}non M., Petit J.-M., 1986, Celes. Mech. 38, 67 
%
\bibitem[\protect\citename{Icke}1982]{Icke}
Icke V., 1982, ApJ, 254, 517
%
\bibitem[\protect\citename{Ida}1990]{I90}
Ida S., 1990, Icarus, 88, 129 
%
\bibitem[\protect\citename{Ida et al.\ }1993]{IKM93}
Ida S., Kokubo E., Makino J. 1993, MNRAS, 263, 875 (IKM93)
%
\bibitem[\protect\citename{Ida \& Makino}1992a]{IM92}
Ida S., Makino J., 1992a, Icarus, 96, 107
%
\bibitem[\protect\citename{Jenkins \& Binney}1990]{JB90}
Jenkins A., Binney J., 1990, MNRAS, 245, 305
%
\bibitem[\protect\citename{Kokubo \& Ida}1992]{KI92}
Kokubo E., Ida S., 1992, PASJ, 44, 601
%
\bibitem[\protect\citename{Lacey}1984]{L84}
Lacey C., 1984, MNRAS, 208, 687 (L84)
%
\bibitem[\protect\citename{Lacey}1991]{L91}
Lacey C., 1991, in Sundelius B., ed., Dynamics of Disk Galaxies.
Dept. of Astronomy/Astrophysics, G\"{o}tenberg Univ., Sweden, P.257
%
\bibitem[\protect\citename{Lissauer \& Stewart}1993]{LS93}
Lissauer J. J., Stewart G. R., 1993, Protostars and Planets III, 
Univ. of Arizona Press, Tucson, P.1061 
%
\bibitem[\protect\citename{Makino}1991]{M91}
Makino J., 1991, PASJ, 43, 859
%
\bibitem[\protect\citename{Makino \& Aarseth}1992]{MA91}
Makino J., Aarseth S. J., 1992, PASJ, 44, 141
%
\bibitem[\protect\citename{Makino et al.\ }1993]{M93}
Makino J., Kokubo E., Taiji M., 1993, PASJ, 45, 349
%
\bibitem[\protect\citename{Makino et al.\ }1997]{M97}
Makino J., Taiji M., Ebisuzaki T., Sugimoto D., 1997, ApJ, 480, 432
%
\bibitem[\protect\citename{Ohtsuki}1998]{O98}
Ohtsuki K., 1998, Icarus, in press
%
\bibitem[\protect\citename{Petit \& H\'{e}non}1986]{PH86}
Petit J.-M., H\'{e}non M., 1986, Icarus, 66, 536
%
\bibitem[\protect\citename{Press et al.\ }1986]{nr}
Press W. H., Flannery B. P., Teukolsky S. A., Vetterling W. T., 1986,
 Numerical Recipes. Cambridge Univ. Press, London/New York.
%
\bibitem[\protect\citename{Spitzer \& Schwarzschild}1953]{SS53}
Spitzer L., Schwarzschild M., 1953, ApJ, 118, 106
%
\bibitem[\protect\citename{Stewart \& Wetherill}1988]{SW88}
Stewart G. R., Wetherill G. W., 1988, Icarus, 74, 542
%
\bibitem[\protect\citename{Stewart \& Ida}1998]{SI98}
Stewart G. R., Ida S., 1998, Icarus, in press
%
\bibitem[\protect\citename{Tanaka \& Ida}1996]{TI96}
Tanaka H., Ida S., 1996, Icarus, 120, 371
%
\bibitem[\protect\citename{Villumsen}1985]{V85}
Villumsen J. V., 1985, ApJ, 290, 75
%
\bibitem[\protect\citename{Wielen}1977]{W77}
Wielen R., 1977, A\&A, 60, 263
%
\end{thebibliography}
\end{document}